\pdfoutput=1
\documentclass[12pt,a4paper]{article}
\usepackage[utf8]{inputenc}
\usepackage{a4wide}
\usepackage{amsmath,amssymb,amsfonts}
\usepackage[square,numbers,sort&compress]{natbib}
\usepackage{color}
\usepackage[colorlinks,allcolors=blue]{hyperref}

\newcommand{\beq}{\begin{eqnarray}}
\newcommand{\eeq}{\end{eqnarray}}
\newcommand{\non}{\nonumber\\}

\newcommand{\SU}{{\rm SU}}
\newcommand{\Og}{{\rm O}}
\newcommand{\SO}{{\rm SO}}
\newcommand{\Sp}{{\rm Sp}}
\newcommand{\p}{\partial}
\renewcommand{\i}{\mathrm{i}}
\renewcommand{\d}{\mathop{}\!\mathrm{d}}
\newcommand{\D}{\mathcal{D}}

\DeclareMathOperator{\PE}{PE}
\newcommand{\rmH}{\mathrm{H}}
\DeclareMathOperator{\diag}{diag}

\newtheorem{conjecture}{Conjecture}

\begin{document}
\begin{titlepage}
  \begin{flushright}
    LU TP 22-65\\
    May 2023
  \end{flushright}
\vskip3cm
\begin{center}
  {\Large\bf Hilbert series and higher-order Lagrangians\\[2mm] for the $\Og(N)$ model}\\[2cm]
  {\bf Johan Bijnens$^{1a}$, Sven Bjarke Gudnason$^{2b}$, Jiahui Yu$^{2c}$, Tiantian Zhang$^{2d}$}\\[2cm]
{$^1$ Department of Astronomy and Theoretical Physics, Lund University,
  Box 43, SE 221-00 Lund, Sweden}\\[0.5cm]
{$^2$ Institute of Contemporary Mathematics, School of Mathematics and Statistics, Henan University, Kaifeng, Henan 475004, P. R. China}
\end{center}
\vfill
\begin{abstract}
We compare the Hilbert series approach with explicit constructions of
higher-order Lagrangians for the $\Og(N)$ nonlinear sigma model. We
use the Hilbert series to find the number and type of operators up to
mass dimension 16, for spacetime dimension $D$ up to 12 and $N$ up to
12, and further classify the operators into spacetime parity and
parity of the internal symmetry group $\Og(N)$. The explicit
construction of operators is done up to mass dimension 12 for both
parities even and dimension 10 for the other three cases. The results
of the two methods are in full agreement. This provides evidence for
the Hilbert series conjecture regarding co-closed but not co-exact
$k$-forms, which takes into account the integration-by-parts
relations.  
\end{abstract}
\vfill
\rule{10cm}{0.5pt}\\
{$^a$\tt johan.bijnens@hep.lu.se}, {$^b$\tt gudnason@henu.edu.cn}, {$^c$\tt yujiahui@henu.edu.cn},\\ {$^d$\tt tt@henu.edu.cn} 
\end{titlepage}

\section{Introduction}

Effective field theories (EFTs) mark the modern viewpoint on quantum field theories, with especially important examples being chiral perturbation theory (ChPT) \cite{Weinberg:1968de,Weinberg:1978kz,Gasser:1983yg,Gasser:1984gg} and the standard model effective field theory (SMEFT) \cite{Weinberg:1979sa,Buchmuller:1985jz,Grzadkowski:2010es}.
The recipe of writing down the EFT is simply to use the symmetries of the problem to write down all possible operators and then order them e.g.~by mass dimension.
Unlike a renormalizable quantum field theory, the EFT is valid only up to some intrinsic scale, be it $\Lambda_{\rm QCD}$ or $\Lambda_{\rm Planck}$, etc.
An EFT is thus parametrized by a number of low-energy constants (LECs) -- how many depends on the order of the operators being taken into account.
The rationale being that the higher the order of the operator, the smaller the contribution to a physical observable in physical computations, which is a simple consequence of all momentum scales in the processes corresponding to the operator, being much smaller than the intrinsic (cut off) scale of the EFT.

Although the precision of a computation is expected to become more precise by going to higher orders, the number of free parameters (operator coefficients) becomes quickly very large, and is expected to grow exponentially. 
A simple question to ask is how many operators exist, compatible with the symmetries of the system, at a given order.
This counting should take into account that one is allowed to integrate by parts (IBP) and perform field redefinitions, since physics is independent of what variables (fields) one chooses to use for the computations. The field redefinitions in an EFT can be shown to be equivalent to using the lowest-order equation of motion (EOM) to reduce operators in the Lagrangian \cite{Scherer:1994wi,Bijnens:1999sh,Grosse-Knetter:1993tae} -- a procedure not allowed in theories that are not EFTs.
Taking the IBP and lowest-order EOM into account, yields the number of independent operators in a minimal basis, which is useful for phenomenology. The basis, of course, is not unique and the choice of which operators to keep and which to eliminate, makes the process of finding the minimal basis at higher orders a formidable task.
In case of ChPT, due to large efforts in the phenomenology community minimal bases for the chiral Lagrangian have been found up to order $\mathcal{O}((\p/\Lambda_{\rm QCD})^8)$ (or simply $p^8$) \cite{Gasser:1983yg,Gasser:1984gg,Fearing:1994ga,Bijnens:1999sh,Bijnens:2018lez} in the normal and to order $p^6$ in the anomalous sector \cite{Bijnens:2001bb,Ebertshauser:2001nj}.
Due to the complexity of the problem in the case of direct computations, the fact that there is no unique basis of operators and there are many relations making different choices of basis equivalent has been a source of confusion in the literature.
Therefore, having a systematic method that with certainty allows one to find the dimension of the basis, i.e.~the number of operators, at a given order in the EFT, provides a crucial check on direct computations.

The Hilbert series method is exactly one such method, that makes it possible to construct all possible operators with a given symmetry and integration using appropriate Haar measures then picks out the invariants -- e.g.~Lorentz and gauge invariants.
In the context of quantum field theory, this method was first used in supersymmetric gauge theories \cite{Benvenuti:2006qr,Feng:2007ur,Gray:2008yu}.
Shortly after, the method has been used to count flavor invariant \cite{Jenkins:2009dy,Hanany:2010vu} and then generalized to EFTs as a powerful toolbox for phenomenologists \cite{Lehman:2015via,Henning:2015daa,Lehman:2015coa,Henning:2015alf,Henning:2017fpj,Ruhdorfer:2019qmk}.
The Hilbert series is a systematic method that not only gives the dimension of the minimal basis of operators (the number of operators), but also the form of the independent operators by means of a graded Hilbert series, albeit without the explicit tensor contractions.
The way to contract the indices amongst a type of operators is not unique, but the different possibilities are related by IBP identities.
An introduction to the method can be found in \cite{Dujava:2022vqz}.

In phenomenology, the discrete symmetries known as parity, charge conjugation and time reversal symmetry, are useful for several reasons. The CPT theorem states that any Lorentz invariant, local and unitary theory is CPT invariant \cite{Zee:2003mt}, which means that breaking of CP is equivalent to breaking of time-reversal symmetry.
Since CP (meaning charge-even and parity-even or charge-odd and parity-odd) is very rarely broken in nature, and because it is envisioned as the source of the breaking of the symmetry between particles versus anti-particles in the early universe \cite{Sakharov:1967dj}, the operators giving rise to CP-violation are important to identify.
In the standard model (SM), the only sources of CP-violation are the $\theta$ angle in QCD, as well as two angles in the CKM and PMNS matrices, where the latter two matrices are mass matrices for the quarks and neutrinos, respectively.
In the Hilbert series method, the parity-even versus parity-odd operators can be found by folding the representations of the Lorentz group \cite{Henning:2017fpj,Graf:2020yxt}.
The technique is analogous for the charge-even versus charge-odd operators, where folding of the representation of the internal symmetry group is performed instead.
Having identified the operators according to these four possibilities, one can readily find CP-even and CP-odd combinations \cite{Graf:2020yxt}.

In this paper, we consider the $\Og(N)$ nonlinear sigma model, which has nonlinearly realized symmetry and spontaneous symmetry breaking of the symmetry group from $\Og(N)$ to $\Og(N-1)$. The Hilbert series method is so far only developed for linearly realized symmetries; however, the nonlinear symmetry can be converted to a linear symmetry using a trick of Callan-Coleman-Wess-Zumino (CCWZ) by using the variables in the coset of the symmetry breaking -- in this case in the coset $\Og(N)/\Og(N-1)$, which transform only with respect to the unbroken group $\Og(N-1)$, but does so linearly \cite{Henning:2017fpj}.
We apply the Hilbert series method, using the CCWZ trick to the $\Og(N)$ nonlinear sigma model, for the first time, and construct the numbers and forms of the operators up to dimension 16 in mass dimension, for up to $D=12$ spacetime dimensions and up to $N=12$, and further classify the operators into four types, according to the parity and "charge" (or rather parity of the $\Og(N)$ group).
We furthermore construct all the operators explicitly using direct methods in field theory and verify agreement between the results. Due to the complexity of the computations using direct methods, we have constructed the operators explicitly only up to dimension 12 for the simplest type of operators and only up to dimension 10 for the remaining types. 
Although the Hilbert series method is in some sense more systematic than the explicit method and it is also more efficient, computationally, the explicit construction method has some advantages, like for instance that the simplest type of operators -- parity-even and charge-even operators -- saturate in numbers when both the spacetime dimension and $N$ are increased.
For the other types of operators, it is also clear that only certain combinations of $D$, $N$ and the dimension $n_d$ of the operator can give a nonvanishing result.
These results agree with the Hilbert series method, but the extraction of such statements in the latter method is yet somewhat obscure. Another advantage of the direct construction method is that one has a fully explicit operator basis.

Alternative methods to obtain Lagrangians come from so-called amplitude methods or different constructions using group theory, see e.g. \cite{Kampf:2013vha,Cheung:2014dqa,Bijnens:2019eze,Fonseca:2019yya,dai:2020cpk,Kampf:2021jvf}, however the work in this area has been restricted to the $\SU(N)$ case, not the $\Og(N)$ model we consider here.

The paper is organized as follows.
In Sec.~\ref{sec:Hilbert} we introduce the Hilbert series method and discuss the modifications of the method specific to the $\Og(N)$ nonlinear sigma model.
In Sec.~\ref{sec:explicit}, we compute the operators explicitly using direct methods in field theory and discuss the explicit relations and constraints utilized there.
Finally, we conclude the paper with a discussion in Sec.~\ref{sec:conclusions}.
The appendices contain 
details about the group characters and representations in App.~\ref{app:characters} and \ref{app:reps}, respectively. 
The Hilbert series results on the total number of operators are delegated to App.~\ref{app:tables} and the actual Hilbert series to App.~\ref{app:explicit_Hilbert}. The Hilbert series contains also information about the number of terms with a given minimal number of fields occurring in the term.
The Hilbert series results are included in the supplementary material file \texttt{Lagrangianshilbertseries.txt} in a machine readable format. Similarly the results from the explicit constructions are given in the supplementary files \texttt{Lagrangianfulltype$i$.txt} for $i=1,\ldots,4$.

\section{Hilbert series}\label{sec:Hilbert}

The Hilbert series for a theory with Lorentz and an internal unbroken
symmetry $H$, is constructed as follows \cite{Henning:2017fpj}
\begin{align}
  \rmH(u,p) &= \rmH_0(u,p) + \Delta\rmH(u,p),\label{eq:H}\\
  \rmH_0(u,p) &= \int\d\mu_H(y)\int\d\mu_{\SO(D)}(x)\frac{Z(u,p,x,y)}{P_+(p,x)},
\end{align}
where $\rmH_0$ is the Hilbert series for all possible $H$ and $\SO(D)$
invariant operators, with the lowest-order equations of motion (EOM) taken into account.
$\Delta\rmH$ on the other hand is the finite series due to the existence
of co-closed but not co-exact forms, that is conjectured only to 
contribute to operators of dimension 2 and $D$.
The addition of the latter takes into account the integrations-by-parts constraints.
$u$ is a (mass) dimension-1 field, to be defined below, $p$ is the
momentum (or derivative operator) and finally $x$ ($y$) are the
coordinates parametrizing the maximal torus of $\SO(D)$ ($H$). 

$Z$ is the generating function of all possible operators in the theory
at hand and the two integrations over the Haar measure of $H$ and the
Lorentz group, respectively, select out the group invariants; that
is, the Lorentz and $H$ invariant operators.

The division by $P_+(p,x)$ -- the momentum generating function -- in the
integrand simply mods out by the overall momentum; that is, it ensures
that momentum conservation (an IBP) is taken into account. 

We now need to write down the generating function for the theory at
hand.
Since the symmetries are not linearly realized in a sigma model, the
necessary but neat trick is to rewrite the theory as a linearly
realized symmetry on a Maurer-Cartan form, due to Callan-Coleman-Wess-Zumino (CCWZ)
\cite{Coleman:1969sm,Callan:1969sn}.

We consider the sigma model type of theory with a spontaneous symmetry
breaking from $G\to H\subset G$ with $X^i\in\mathfrak{g}/\mathfrak{h}$
being the broken generators parametrizing Nambu-Goldstone bosons and
$T^a\in\mathfrak{h}$ being the unbroken generators.
The nonlinear field is
\beq
\xi = \exp\left(\frac{\i\phi^iX^i}{f_\phi}\right),
\label{eq:xi}
\eeq
which transforms as
\beq
\xi \to \xi' = g\xi h^{-1}(g,\xi),\qquad
g\in G,\qquad
h\in H.
\eeq
Notice that although $g$ is a global transformation, $h$ is local in
that it depends on $\xi$, thereby realizing the nonlinear symmetry.
The left-invariant Maurer-Cartan form is written as
\beq
w = w_\mu\d x^\mu
= \xi^{-1}\p_\mu\xi\d x^\mu
= u_\mu^iX^i\d x^\mu + v_\mu^a T^a\d x^\mu
= u_\mu\d x^\mu + v_\mu\d x^\mu
\in\Lambda^1,
\label{eq:w}
\eeq
where $u_\mu\in\mathfrak{g}/\mathfrak{h}$ lives in the coset space
and $v_\mu\in\mathfrak{h}$ belongs to the stabilizer or unbroken
algebra.
$w$ transforms only according to the unbroken group, $H$:
\beq
w\to w' = h(w + \d)h^{-1}
= h u h^{-1} + h(v + \d)h^{-1}, \qquad h\in H,
\eeq
where the grouping in the last equality means that $u$ transforms
homogeneously whereas $v$ transforms inhomogeneously.
This can be seen from the fact that $u\in\mathfrak{g}/\mathfrak{h}$,
whereas both $v\in\mathfrak{h}$ and $h d h^{-1}\in\mathfrak{h}$.

The building blocks of the Lagrangian and their transformation rules
are
\begin{align}
\xi&\to g\xi h^{-1}, \\
u&\to h u h^{-1}, \\
\D:=\d + v&\to h \D h^{-1}, \\
F:=\D\wedge\D&\to h F h^{-1},\quad g\in G,\; h\in H,
\end{align}
which are 0-, 1-, 1- and 2-forms, respectively;
clearly, only $\xi$ transforms under $g$.
Now, (covariant) derivatives of $\xi$ can be traded for $u$, and $\xi$:
\beq
\D\xi = \d\xi - \xi v = \xi u,\qquad
\D\xi^{-1} = d\xi^{-1} + v\xi^{-1} = -u\xi^{-1},
\eeq
where we have used the definition $w=\xi^{-1}\d\xi=u+v$ and that
$\d(\xi^{-1}\xi)=0$ implying that $\d\xi^{-1}\xi=-\xi^{-1}\d\xi$.
Since the Lagrangian is $G$-invariant and only $\xi$ transforms under
$G$, the only way to obtain $G$-invariants is by multiplying the $\xi$
terms with $\xi^{-1}$, but since $\xi^{-1}\xi=1$ then $\xi$ simply
drops out of the Lagrangian.
The remaining building blocks are now $u$, $\D$ and $F$ which only
transform under $H$.
Imposing $H$-invariance is then equivalent to the original
$G$-invariance.

We can further eliminate $F$ by the following argument.
Since $w$ is a Maurer-Cartan 1-form, it has vanishing curvature (field
strength):
\begin{align}
0
&= \d w + w\wedge w\non
&= \d(u + v) + (u + v)\wedge(u + v)\non
&= \D u + u\wedge u + F,
\end{align}
and clearly $F$ can be traded for $\D u$ and $u\wedge u$.
Decomposition onto the algebra makes the statement even stronger.
$F=\D\wedge\D\in\mathfrak{h}$ is in the unbroken algebra whereas
$\D u\in\mathfrak{g}/\mathfrak{h}$ lives in the coset space (as
$[T^a,X^i]\in\mathfrak{g}/\mathfrak{h}$).
In fact, generically speaking $u\wedge u\in\mathfrak{g}$ and so has
components in both the coset space and the unbroken algebra
$u\wedge u=u\wedge u|_{\mathfrak{h}} + u\wedge u|_{\mathfrak{g}/\mathfrak{h}}$.
However, for symmetric spaces $u\wedge u|_{\mathfrak{g}/\mathfrak{h}}=0$.
In particular,\footnote{For $SO(3)/SO(2)=S^2\simeq\mathbb{C}P^1$ is a complex projective space, which is K\"ahler and advantageous for calculations, see e.g.~\cite{Gudnason:2021gcp}; this is not the case for higher $N$ in the $\SO(N)$ model. }
\begin{align}
G/H&=\SO(N)/\SO(N-1)=S^{N-1},\non
G/H&=\SO(N)/({\rm S}[\Og(N-1)\times\Og(1)])=\mathbb{R}P^{N-1},
\end{align}
are symmetric spaces for which $u\wedge u|_{\mathfrak{g}/\mathfrak{h}}=0$. 
In any case, the conclusion is that both $F$ and $\D u$ can be traded
for a polynomial in $u$; 
but in our case of a symmetric coset space, $\D u=0$.
This means that the building blocks can now be reduced to $u$ and any
symmetric traceless single particle module
\beq
R_u =
\begin{pmatrix}
  u_{\mu_1}\\
  \D_{\{\mu_1}u_{\mu_2\}}\\
  \D_{\{\mu_1}\D_{\mu_2}u_{\mu_3\}}\\
  \vdots
\end{pmatrix}.
\label{eq:Rmu}
\eeq
In particular, the equation of motion $\D_\mu u_\mu=0$ is also taken
into account automatically by the traceless condition.

The generating function for the single particle module $R_u$ of the
operator space $\oplus_{n=0}^\infty{\rm sym}^n(R_u)$ is given by
\beq
Z(u,p,x,y)
= \sum_{n=0}^\infty u^n\chi_{{\rm sym}^n(R_u)}(p,x,y)
= \PE[u\chi_u(p,x,y)],
\eeq
where $u$ is now a weight label, $\PE$ is the plethystic exponential and
the character $\chi_u$ being the infinite sum of $p$-weighted $\SO(D)$
characters times $H$ characters at dimension $k$:
\begin{align}
\chi_u(p,x,y)
&=\sum_{k=0}^\infty \chi_{u,k}(p,x,y)\non
&=\sum_{k=0}^\infty p^{k+1}\chi_{(k+1,0,\cdots,0)}(x)\chi_{H,u}(y)\non
&=\left[\sum_{k=0}^\infty p^k\chi_{(k,0,\cdots,0)}(x) - 1\right]\chi_{H,u}(y)\non
&=\left[\sum_{k=0}^\infty p^k\left(\chi_{{\rm sym}^k(\square)}(x) - p^2\chi_{{\rm sym}^k(\square)}(x)\right) - 1\right]\chi_{H,u}(y)\non
&=\left[(1-p^2)P_+(p,x) - 1\right]\chi_{H,u}(y),
\label{eq:chi_u}
\end{align}
where
\beq
P_+(p,x) = \sum_{k=0}p^k\chi_{{\rm sym}^k(\square)}(x)
=\begin{cases}
\prod_{i=1}^r\frac{1}{(1-p x_i)(1-p/x_i)}, & d=2r,\\
\frac{1}{1-p}\prod_{i=1}^r\frac{1}{(1-p x_i)(1-p/x_i)}, & d=2r+1,
\end{cases}
\label{eq:P+}
\eeq
is the momentum generating function in $D$ dimensions.
The reason for the $-p^2$ term is due to conformal representation
theory and that the symmetric traceless tensor product of the
fundamental representation furnishes a short representation of the
conformal group $\SO(D,2)$ for a scalar field (the
$\chi_{(k,0,\cdots,0)}(x)$ term).
This is exactly the EOM (for a scalar field) being removed in all
possible combinations from all possible operators.
The $-1$ term, on the other hand, removes the scalar component from
the single particle module \eqref{eq:Rmu}, since the lowest component
field is a vector field.\footnote{The single particle module for a scalar field
is related to that of a vector field by removing the first component
(the scalar component); this is the $-1$. The truncated single
particle module, however, does not correspond to a field that
transforms under the conformal group. }
$\chi_{H,u}(y)$ is the group character for the internal unbroken symmetry group $H$.
Since we are considering $G=\SO(N)$ in the vector representation and spontaneous symmetry breaking implies $H=\SO(N-1)$, also in the vector representation, the group characters are given by
\beq
\chi_{H,u}(y) = 
\begin{cases}
  \sum_{i=1}^r\left(x_i + x_i^{-1}\right), & N-1=2r,\\
  1+\sum_{i=1}^r\left(x_i + x_i^{-1}\right), & N-1=2r+1.
\end{cases}
\label{eq:chi_Hu}
\eeq
The $\PE$ is simply
\beq
Z(u,p,x,y) = \PE[u\chi_u(p,x,y)]
= \exp\left(\sum_{r=1}^\infty\frac1r u^r\chi_u(p^r,x^r,y^r)\right).
\eeq
The Haar measure for $\SO(D)$ is a product over all roots of the algebra.
However, since the integrals are over Weyl-invariant quantities, we can use the Haar measure restricted to the product over positive roots
\begin{align}
\d\mu_{\SO(2r)} &= \prod_{k=1}^r\frac{\d x_k}{2\pi\i x_k}
\!\prod_{1\leq i<j\leq r}\!(1-x_ix_j)\left(1-\frac{x_i}{x_j}\right),\\
\d\mu_{\SO(2r+1)} &=\prod_{k=1}^r\frac{\d x_k}{2\pi\i x_k}
(1-x_k)\!\prod_{1\leq i<j\leq r}\!(1-x_ix_j)\left(1-\frac{x_i}{x_j}\right).
\end{align}

\subsection{Parity and intrinsic parity}\label{sec:HS_parity}

We will now consider imposing spacetime parity and internal parity symmetries, following \cite{Henning:2017fpj,Graf:2020yxt}.
By spacetime parity symmetry, we intend an overall sign flip, $x\to-x$ for odd $D$ dimensions and a sign flip of the last component of the coordinate vector, $(x^1,\cdots,x^{D-1},x^D)\to(x^1,\cdots,x^{D-1},-x^D)$ for even $D$ dimensions.
We note that this spacetime parity symmetry is different from the conventional (spatial) parity symmetry in 3+1 dimensions, where $(x^1,x^2,x^3)\to(-x^1,-x^2,-x^3)$.
It is, however, related to the Euclidean spacetime parity symmetry considered here, by a $\pi$ rotation in the $(x^1,x^2)$-plane. 

The internal parity is essentially the same, albeit for the $H=\SO(N-1)$ group.
Similarly, for $N$ even the overall sign of the field $u$ is simply flipped, whereas for $N$ odd (corresponding to $N-1$ even), only the last component of the $N-1$ vector is flipped.

Starting with the spacetime parity, we first notice that $\SO(D)$ does
not contain the parity symmetry, it is however easily included by
changing the Lorentz symmetry for even $D$ from $\SO(D)$ to 
$\Og(2r)=\SO(2r)\ltimes\mathbb{Z}_2=(\Og_+(2r),\Og_-(2r))$, with
$\SO(2r)=\Og_+(2r)$, $D=2r$ and $\ltimes$ is the semi-direct product,
whereas for odd $D$, the semi-direct product should be replaced by the
direct product: $\Og(2r+1)=\SO(2r+1)\times\mathbb{Z}_2=(\Og_+(2r+1),\Og_-(2r+1))$,
with $\SO(2r+1)=\Og_+(2r+1)$, $D=2r+1$ and $\times$ is the direct product.
The parity-even case is calculated as
\beq
\rmH^{P\text{-even}} = \frac12(\rmH^{P^+} + \rmH^{P^-}),
\label{eq:Heven}
\eeq
whereas the parity odd case is the compliment
\beq
\rmH^{P\text{-odd}} = \frac12(\rmH^{P^+} - \rmH^{P^-}).
\eeq
The first Hilbert series, $H^{P^+}$ in the above expressions is the one
calculated in the previous section and is the total Hilbert series.
The parity symmetry can be viewed as splitting the total series into
parity-even and parity-odd operators.

The $P^-$ corresponds to switching the Lorentz group $\SO(D)$ for
$\Og_-(D)$, which for the Haar measure implies
\beq
\d\mu_{\Og_-(D)} =
\begin{cases}
  \d\mu_{\Sp(2r-2)}, & D=2r,\\
  \d\mu_{\SO(2r+1)}, & D=2r+1,
\end{cases}
\eeq
which is a result that can be obtained by folding
\cite{Henning:2017fpj}.
The Haar measure for $\Sp(D-2)$ for Weyl-invariant quantities (taking into account only the product over positive roots), is given by
\beq
\d\mu_{\Sp(2r-2)} = 
\prod_{k=1}^{r-1}\frac{\d x_k}{2\pi\i x_k}
\prod_{k=1}^{r-1}(1-x_k^2)
\prod_{1\leq i\leq j\leq r-1}(1 - x_ix_j)\left(1-\frac{x_i}{x_j}\right).
\eeq
The momentum generating function for $\Og_-(D)$ reads
\begin{align}
P_{-}(p,x) &= 
\begin{cases}
  \frac{1}{1-p^2}P_+^{2r-2}(p,\tilde{x}), & D=2r,\qquad \tilde{x}=(x_1,x_2,\cdots,x_{r-1})\\
  P_+(-p,x), & D=2r+1,
\end{cases}\non
&=
\begin{cases}
  \frac{1}{1-p^2}\prod_{i=1}^{r-1}\frac{1}{(1-px_i)(1-p/x_i)}, & D=2r,\\
  \frac{1}{1+p}\prod_{i=1}^r\frac{1}{(1+px_i)(1+p/x_i)}, & D=2r+1,
\end{cases}
\label{eq:P_-}
\end{align}
and finally the vector property of the field $u$ yields an overall
sign flip for the group character in even dimensions
\beq
\chi_u^{P^-}(p,x,y) =
\begin{cases}
  -\left[(1-p^2)P_-(p,x) - 1\right]\chi_{H,u}(y), & D=2r,\\
  \left[(1-p^2)P_-(p,x) - 1\right]\chi_{H,u}(y), & D=2r+1.\\
\end{cases}
\label{eq:chi_u_-}
\eeq
An extra subtlety happens in even dimensions for even powers of the
characters in the plethystic exponential ($\PE$); this can be dealt
with as follows
\begin{align}
Z^{P^-}(u,p,x,y) =
\exp\bigg(&\sum_{r=1}^\infty\frac{1}{2r}u^{2r}\chi_u(p^{2r},\bar{x}^{2r},y^{2r})\non&
+\sum_{r=1}^\infty\frac{1}{2r-1}u^{2r-1}\chi_u^{P^-}(p^{2r-1},\tilde{x}^{2r-1},y^{2r-1})
\bigg),
\end{align}
for $D=2r$, where
\beq
x = (x_1,x_2,\cdots,x_r),\qquad
\tilde{x} = (x_1,x_2,\cdots,x_{r-1}),\qquad
\bar{x} = (x_1,x_2,\cdots,x_{r-1},1).
\eeq
The above results hold for generic $D$ with $r>1$.
When $r=1$, the rank is so small that the last component of the
highest weight is nonvanishing for a vector representation, which
leads to the exception in even dimensions, i.e.~$D=2$, when folding
\beq
\chi_u^{P^+} = \chi_u(p,1,y), \qquad
\chi_u^{P^-} = 0, \qquad
\d\mu = 1.
\eeq

We can now contemplate including also intrinsic parity and we will use the symbol $C^\pm$ for representing the positive and negative chambers of the orthogonal group.
$C^-$ corresponds now to switching $H=\SO(N-1)$ for $O_-(N-1)$, for
which the Haar measures are
\beq
\d\mu_{O_-(N-1)} = 
\begin{cases}
  \d\mu_{\Sp(2r-2)}, & N-1=2r,\\
  \d\mu_{\SO(2r+1)}, & N-1=2r+1.
\end{cases}
\eeq
For even $N$, $N-1$ is odd and the $C^-$ partition function is simply
given by
\beq
Z^{C^-}(u,p,x,y) = Z(-u,p,x,y),
\eeq
whereas for odd $N$, $N-1$ is even and the folding of the $\SO(N-1)$
algebra is again done.
In that case, analogously to the case of the Lorentz group, the
characters become
\begin{align}
\chi_{\SO(2r),u}^{C^-}(y) 
= \chi_{\Sp(2r-2),u}(\tilde{y})
= \sum_{i=1}^{r-1}\left(y_i + y_i^{-1}\right),
\end{align}
and the partition function again breaks up into odd and even powers in
the $\PE$ as
\begin{align}
Z^{C^-}(u,p,x,y) =
\exp\bigg(&\sum_{r=1}^\infty\frac{1}{2r}u^{2r}\chi_u(p^{2r},x^{2r},\bar{y}^{2r})\non&
+\sum_{r=1}^\infty\frac{1}{2r-1}u^{2r-1}\chi_u^{C^-}(p^{2r-1},x^{2r-1},\tilde{y}^{2r-1})
\bigg),
\end{align}
where
\beq
y = (y_1,y_2,\cdots,y_r),\qquad
\tilde{y} &= (y_1,y_2,\cdots,y_{r-1}),\qquad
\bar{y} &= (y_1,y_2,\cdots,y_{r-1},1).
\eeq
Now the Hilbert series for even-internal parity operators is given by
\beq
\rmH^{C\text{-even}} = \frac12(\rmH^{C^+} + \rmH^{C^-}),
\eeq
whereas the odd-internal parity case is the compliment
\beq
\rmH^{C\text{-odd}} = \frac12(\rmH^{C^+} - \rmH^{C^-}).
\eeq
The combination of $P^+C^+$, $P^-C^+$, $P^+C^-$, $P^-C^-$ is thus
straightforward and in particular, the Hilbert series of the 4 types
are given as
\begin{align}
\label{eq:type1}
\text{type 1}: \rmH^{P\text{-even},C\text{-even}} 
&=\frac14\left(\rmH^{P^+C^+} + \rmH^{P^+C^-} + \rmH^{P^-C^+} + \rmH^{P^-C^-}\right),\\
\label{eq:type2}
\text{type 2}: \rmH^{P\text{-odd},C\text{-even}} 
&=\frac14\left(\rmH^{P^+C^+} + \rmH^{P^+C^-} - \rmH^{P^-C^+} - \rmH^{P^-C^-}\right),\\
\label{eq:type3}
\text{type 3}: \rmH^{P\text{-even},C\text{-odd}} 
&=\frac14\left(\rmH^{P^+C^+} - \rmH^{P^+C^-} + \rmH^{P^-C^+} - \rmH^{P^-C^-}\right),\\
\label{eq:type4}
\text{type 4}: \rmH^{P\text{-odd},C\text{-odd}} 
&=\frac14\left(\rmH^{P^+C^+} - \rmH^{P^+C^-} - \rmH^{P^-C^+} + \rmH^{P^-C^-}\right).
\end{align}
A nontrivial check for the representation theory to be correct, is
that all coefficients of all four types of Hilbert series must be
positive integers.
Notice that although the coefficients of $\rmH^{P^+C^-}$, $\rmH^{P^-C^+}$ and $\rmH^{P^-C^-}$ also must be integer, they are not necessarily positive.

\subsection{The exceptional Hilbert series}\label{sec:HS_exceptional}

We still have to address the operators that are not taken correctly into account by $\rmH_0$ and are represented in Eq.~\eqref{eq:H} by $\Delta\rmH$.
Let us briefly review the argument \cite{Henning:2017fpj}.
The number of operators that we would like to count is
\beq
\#({\rm ops}) = \#(\Lambda^0) - \#(\Lambda_{\textrm{co-exact}}^0),
\eeq
where the co-exact 0-forms are all the 1-forms that are not co-closed:
\beq
\#(\Lambda_{\textrm{co-exact}}^0) = \#(\Lambda^1) - \#(\Lambda_{\textrm{co-closed, but not co-exact}}^1) - \#(\Lambda_{\textrm{co-exact}}^1),
\eeq
where $\Lambda^k$ are $k$-forms and $\#(X)$ means the number of the object $X$.
Iterating the forms up to co-exact $(d-1)$-forms, we have
\beq
\#({\rm ops}) =
\sum_{k=0}^d(-1)^k\#(\Lambda^k)
+\sum_{k=1}^d(-1)^{k+1}\#(\Lambda_{\textrm{co-closed, but not co-exact}}^k),
\eeq
where the first sum is $\rmH_0$ in \eqref{eq:H} and the last is $\Delta\rmH$. 

$\Delta\rmH$ consists of operators that are co-closed, but not co-exact forms, which is differential geometric language for operators
\beq
\delta\omega = 0,
\eeq
where $\omega\in\Lambda^r$ cannot be written as $\delta\lambda$, for any $\lambda$, and $\delta$ is the coderivative $\delta=(-1)^{D(r+1)+1}*\d*$, defined in terms of the exterior derivative and the Hodge star operation.
Clearly $\delta^2=0$, since $\d^2=0$ by antisymmetry of differential forms and $\delta^2=*\d**\d*=(-1)^{r(D-r)}*\d^2*=0$, where we have used that $*^2=(-1)^{r(D-r)}$ in $D$-dimensional Euclidean space for an $r$-form.
In components, an $r$-form, $\omega\in\Lambda^r$,
\beq
\omega=\frac{1}{r!}\omega_{\mu_1\mu_2\cdots\mu_r}\d x^{\mu_1}\wedge\d x^{\mu_2}\wedge\cdots\wedge\d x^{\mu_r},
\eeq
acted on by the coderivative
\beq
\delta\omega=(-1)^{r(r-1)-1}\frac{1}{(r-1)!}\p_\nu\omega_{\nu\mu_1\cdots\mu_{r-1}}\d x^{\mu_1}\wedge\cdots\wedge\d x^{\mu_{r-1}},
\eeq
is simply the divergence of the tensor.

We will now present a conjecture, which is an adaptation of that of
Ref.~\cite{Henning:2017fpj} to the $\Og(N)$ model.
In addition to the arguments presented there, we point out that the
Poincar\'e lemma tells us that all co-closed forms are also co-exact;
the exceptions are those cases where the co-exact forms can only be
written using field redefinitions.
We do not have a rigorous proof of there being no further exceptions.

\begin{conjecture}
The only co-closed, but not co-exact forms that will contribute to the Hilbert series in the $\SO(N)$ nonlinear sigma model are given by
\beq
*(\wedge^{N-1}u), \qquad
*1,
\eeq
with contribution to the exceptional Hilbert series $\Delta\rmH$ of Eq.~\eqref{eq:H}:
\beq
\Delta\rmH = (-1)^{D-N}p^{D-N+1}u^{N-1}
+ (-1)^{D+1}p^D.
\eeq
\end{conjecture}
The only possible operators that correspond to the above differential forms, that are $\SO(N-1)$-invariant ($H$-invariant) are given by
\begin{align}
    \epsilon^{a_1a_2\cdots a_{N-1}}
    \epsilon^{\mu_1\mu_2\cdots\mu_{N-1}\mu_{N}\cdots\mu_{D}}
    u_{\mu_1}^{a_1}u_{\mu_2}^{a_2}\cdots u_{\mu_{N-1}}^{a_{N-1}}\d x^{\mu_{N}}\wedge\cdots\wedge\d x^{\mu_{D}},\qquad N \leq D+1,
\end{align}
which we can see is co-closed by the fact that the EOM constraint $\delta u=0$ vanishes.
This form is not co-exact, since there are no form whose divergence yields the antisymmetric product of only $u$s.

First let us note that all co-closed forms are also co-exact by the Poincar\'e lemma on $\mathbb{R}^D$ and by using duality.
That is
\beq
*\d*u = 0, \quad\Rightarrow\quad \d*u=0,
\eeq
and therefore $\mathop*u$ is closed.
By the Poincar\'e lemma on Euclidean space, we can always write 
\beq
\mathop*u = \d(\mathop*\lambda).
\eeq
Taking the Hodge dual of the above, we have that $u=\delta\lambda$ up to a sign.
Although this is always true by de Rham cohomology theory, the obstruction in our case of counting operators, is that we only allow for $\delta u=0$ as a field redefinition (recall that taking into account the lowest-order EOM corresponds to taking into account field redefinitions).
For instance, the (gauge variant) 1-form $u$ can geometrically be written as the divergence of a 2-form $u=\delta\lambda$, but that would require a change of variables and we have defined the Hilbert series counting scheme as counting only $u$s and $\d$s.
The exception is thus that all $r$-forms that do not contain derivative operators, cannot be co-exact.
These are
\beq
\mathop*1,\quad
\mathop*u,\quad
\mathop*(u\wedge u), \quad
\cdots, \quad
\mathop*(u\wedge\cdots\wedge u),
\label{eq:differentialless}
\eeq
with a maximum of $D$ or $N-1$ $u$s.
However, only the terms with $0$ and $N-1$ $u$s are $H$-invariant.
Finally, the volume form is not co-exact, since there exist no $(D+1)$-forms.

\subsection{Results}

We now implement the Hilbert series in a \textsc{Mathematica} notebook as well as in \textsc{FORM} and obtain results in agreement with each other as well as in agreement with the explicit construction method, discussed in the next section.
The total number of operators in a minimal basis of operators of
dimension $n_d$ for the $G=\SO(N)$ nonlinear sigma model in $D$
Euclidean dimensions, are presented for\footnote{The types are defined in Eqs.~\eqref{eq:type1}-\eqref{eq:type4}.} type 1 in tables
\ref{tab:resultstype1} and \ref{tab:resultstype1b}, for type 2 in
tables \ref{tab:resultstype2} and
\ref{tab:resultstype2b}-\ref{tab:resultstype2c}, for type 3 in tables
\ref{tab:resultstype3} and \ref{tab:resultstype3b}-\ref{tab:resultstype3c} and for type 4 in
tables \ref{tab:resultstype4}-\ref{tab:resultstype4b} and \ref{tab:resultstype4c}-\ref{tab:resultstype4e}.
We have calculated the Hilbert series results for $n_d$ up to 16 with $D$ and $N$ up to 12.

An advantage of the Hilbert series method, as can be seen from the tables, is that the results can be obtained for larger operator dimension $n_d$ and larger $N$ and $D$, with respect to running time and memory consumption of a PC, as compared to the explicit construction method.
The implementation of the Hilbert series method is also straightforward, especially if only the total number of operators is required; whereas splitting the series into type 1 through 4 requires a more complicated implementation, as described in sec.~\ref{sec:HS_parity}.

An advantage of the explicit construction method over the Hilbert series method, is that it is more clear which combinations of $n_d$, $D$ and $N$ must vanish for types 2 through 4 and it is clear that the number of operators of type 1 saturates in number for sufficiently large $D$ and sufficiently large $N$. 
From the Hilbert series method, this effect is observed from the results, but is less clear from the formulae. 

The version of the Hilbert series method utilized here is based on differential forms and the exceptional Hilbert series consists by definition of only co-closed, but not co-exact forms. In Ref.~\cite{Graf:2020yxt}, it was conjectured that the only contributions to $\Delta\rmH$ were given by the forms of the type discussed in sec.~\ref{sec:HS_exceptional}.
Although we have not given a proof of this conjecture, we confirm it for all the operators that are computed by both the Hilbert series method and the explicit construction method, which are type 1 operators with $n_d$ up to 12 and types 2 through 4 operators with $n_d$ up to 10.

\section{Explicit construction}\label{sec:explicit}

This is the method traditionally used in the EFT community.

In this section we use a vector rather than the generator notation used in the previous section. For transformations $g\in ({\rm S})\Og(N)$, we introduce a real vector field of size $N$, $\Phi$ (column vector). Spontaneous symmetry breaking is implemented by the requirement $\Phi^T\Phi=1$. Transformation under the full symmetry $G$ is $\Phi\to g\Phi$.

Assuming that we choose $\Phi$ such that its vacuum expectation value is: $\Phi^T=(1,0,\ldots,0)$ a general parametrization of $\Phi$ in terms of an $N-1$ column vector $\phi$ is
\begin{align}
  \label{eq:defPhi}
\Phi = \left(\sqrt{1-\left(f\left(\frac{\phi^T\phi}{F^2}\right)\right)^2\frac{\phi^T\phi}{F^2}},\,f\left(\frac{\phi^T\phi}{F^2}\right)\frac{\phi^T}{F}\right).
\end{align}
$\Phi$ transforms as expected under $({\rm S})\Og(N)$. The function $f(x)$ is real and analytic and satisfies $f(0)=1$. This parametrization has $({\rm S})\Og(N-1)$ as an explicit symmetry via $\phi\to h\phi$ with $h\in ({\rm S})\Og(N-1)$. Varying the choice of $f(x)$ has been used in several papers as a check on explicit calculations,
see e.g.~\cite{Bijnens:2009zi,Bijnens:2010xg,Bijnens:2021hpq}.
The field $\phi$ and the constant $F$ are chosen to have dimension $D/2$, so $\phi$ is conventionally normalized and $\Phi$ is dimensionless. This construction is fully equivalent to the general method of \cite{Coleman:1969sm,Callan:1969sn} used in the previous section. 

We denote the group index here by letters $a,b,c,\ldots$ running from $1$ to $N$ and spacetime or Lorentz indices by $\mu,\nu,\rho,\ldots$ running from $0$ to $D-1$. No other types of indices appear. For group indices we will use an index free notation as much as possible. Lorentz indices we will always indicate explicitly.

Invariants can be produced by contracting indices or by contracting them with a group or Lorentz Levi-Civita tensor. Given that the product of two Levi-Civita tensors with indices of the same type can be rewritten in terms of Kronecker deltas, we need only to consider at most one of each in constructing invariants.
 So there exists four types of terms:
\begin{enumerate}
\item \label{T1} Containing pairs of Lorentz-indices via $\partial_\mu \partial^\mu$ and pairs of $O(N)$ indices via $\Phi^{(1)}_a \Phi^{(2)}_a$; even under parity and intrinsic parity. These are type 1 of Eq.~\eqref{eq:type1}.
\item \label{T2} In addition to the content of the first type, also one $\epsilon^{\mu_1\ldots \mu_{D}}\partial_{\mu_1}\ldots\partial_{\mu_{D}}$; odd under parity, even under intrinsic parity.  These are type 2 of Eq.~\eqref{eq:type2}.
\item \label{T3} In addition to the content of the first type, also one $\epsilon^{a_1\ldots a_N}\Phi^{(1)}_{a_1}\ldots\Phi^{(N)}_{a_N}$; even under parity, odd under intrinsic parity.  These are type 3 of Eq.~\eqref{eq:type3}.
\item \label{T4} In addition to the content of the first type, one of $\epsilon^{\mu_1\ldots \mu_{D}}\partial_{\mu_1}\ldots\partial_{\mu_{D}}$ and one of  $\epsilon^{a_1\ldots a_N}\Phi^{(1)}_{a_1}\ldots\Phi^{(N)}_{a_N}$; odd under parity and intrinsic parity.  These are type 4 of Eq.~\eqref{eq:type4}.
\end{enumerate}
The partial derivatives can act on any object. $\Phi^{(i)}$ indicates an object transforming under $({\rm S})\Og(N)$ as $\Phi$. The four types are distinguished by parity and intrinsic parity as indicated.

In the remainder we know that Lorentz indices are always contracted between one lower and one upper index so we ignore the distinction in the notation. We also introduce the notation\footnote{$f$ is not be confused with function $f$ in the general parametrization of Eq.~\eqref{eq:defPhi}.}
\begin{align}
\partial_{\mu_1}\ldots\partial_{\mu_n}\Phi_a &\equiv \Phi_{a;\mu_1\ldots\mu_n}\,,\\
\Phi_{a;\mu_1\ldots\mu_n}\Phi_{a;\nu_1\ldots\nu_m}&\equiv\Phi^T_{\mu_1\ldots\mu_n}\Phi_{\nu_1\ldots\nu_m}= f_{\mu_1\ldots\mu_n;\nu_1\ldots\nu_m}\,,\\
\epsilon^{a_1\ldots a_N}\Phi_{a_1;\mu_1^{(1)}\ldots}\ldots\Phi_{a_N;\mu_N^{(N)}\ldots}&\equiv
g_{\mu_1^{(1)}\ldots;\ldots;\mu_N^{(N)}\ldots}\,.
\end{align}
This allows us to write Lagrangians without explicit group indices.

The lowest order, two derivatives, Lagrangian is given by
\begin{align}
  \label{eq:L2}
  \mathcal{L}_2 = \frac{F^2}{2}\partial^\mu\Phi^T\partial_\mu\Phi = \frac{F^2}{2} f_{\mu;\mu}.
\end{align}
For the case $D=2$, $N=3$ there exists one more term with two derivatives, i.e.~the pullback of the area form on $S^2$ by $\Phi$:
\begin{align}
\label{eq:CS}
    \mathcal{L}_{\rm topo} &= c_{\rm topo} \epsilon_{\mu\nu}g_{~;\mu;\nu} = c_{\rm topo}\epsilon_{\mu\nu}\epsilon^{abc}\Phi_a \Phi_{b;\mu} \Phi_{c;\nu}\,.
\end{align}
It is a topological term, somewhat similar to Chern-Simons, but different because $\Phi$ is not a spacetime vector, but an $\SO(N)$ vector.
This term is often integrated (with proper normalization by the volume of the 2-sphere) to count the number of lumps or vortices in $S^2$ or $\mathbb{C}P^1$ models \cite{Manton:2004tk}.

Terms in the Lagrangian are related by many things:
\begin{enumerate}
\item \label{R1} Partial derivatives commute.
\item \label{R2} The fields $\Phi$ also commute.
\item \label{R3} $\Phi^T\Phi=1$ has many consequences obtained from derivatives acting on $\Phi^T\Phi=1$.
\item \label{R4} A total derivative does not contribute, i.e.~terms related via partial integration are the same.
\item \label{R5} Field redefinitions respecting the symmetry can be used to remove terms, alternatively the lowest order equation of motion can be used to find which terms can be removed that way.
\item \label{R6} Schouten identities I: In $D$ dimensions there is no fully antisymmetric combination of $D+1$ different Lorentz indices.
\item \label{R7} Schouten identities II: For $O(N)$ there is no fully antisymmetric combination of $N+1$ different flavor indices.
\item \label{R8} For a given value of $N$ there are additional consequences of $\Phi^T\Phi=1$. These are derived in Sec.~\ref{sec:determinant}.
\end{enumerate}
In explicitly constructing Lagrangians it is not easy to know if one has included all possible symmetry constraints, the comparison with the Hilbert series is very useful for knowing the Lagrangian is really minimal.

Field redefinitions: assume a small variation $\Psi$. It should have the same transformations under all symmetries as $\Phi$.
It can be considered general for deriving the lowest order equations of motion
or can be restricted to containing derivatives and factors of $\Phi$ such that it vanishes for $\phi=0$ and has at least one derivative. In that case it be considered as a field redefinition only affecting higher-order Lagrangians.

The transformations
\begin{align}
\label{eq:fieldredef}
  \Phi\to\frac{\Phi+\Psi}{\sqrt{1+2\Phi^T\Psi+\Psi^T\Psi}}\approx \Phi+\Psi-\Phi \Phi^T\Psi+\cdots
\end{align}
are thus the most general field redefinitions consistent with the constraints.

The lowest-order equation of motion (EOM) can be derived from (\ref{eq:L2}) using (\ref{eq:fieldredef}) or using the Euler-Lagrange equations:
\begin{align}
\label{eq:EOM}
  \partial^2 \Phi + \Phi \partial_\mu\Phi^T\partial^\mu\Phi=0.
\end{align}
The EOM of (\ref{eq:EOM}) is an $N$-component equation but applying $\partial^2$ to $\Phi^T\Phi=1$ leads to the relation
\begin{align}
  \Phi^T\left( \partial^2 \Phi + \Phi \partial_\mu\Phi^T\partial^\mu\Phi\right)=0,
\end{align}
so that there only $N-1$ independent equations.

The variation of the LO Lagrangian under (\ref{eq:fieldredef}) is
\begin{align}
  \label{eq:varLOLag}
  \delta\mathcal{L}_2 = \partial_\mu\Phi^T\partial^\mu\Psi-\partial_\mu\Phi^T\partial^\mu\Phi \partial\Phi^T\Psi+\ldots
  = -\left(\partial^2\Phi-\partial_\mu\Phi^T\partial^\mu\Phi\,\Phi\right)^T\Psi+\cdots\,,
\end{align}
where we used $\Phi^T\partial_\mu\Phi=0$, which is a consequence of $\Phi^T\Phi=1$, and partial integration. By choosing $\Psi$ appropriately one can recursively remove higher-order terms that contain the equation of motion. The other terms are of order $\Psi^2$ and higher order than the term shown.

\subsection{Extra relations}
\label{sec:determinant}

Let us look at the $N\times N$ matrix, with rows labeled by an index $k$
referring to which of a set of $N$ Lorentz indices ${\mu_{i}}$ a derivative is taken with, and the column by which of the components of $\Phi$ is used:
\begin{align}
  \label{eq:determinant}
  \det\begin{pmatrix}\partial_{\mu_1}\Phi_1 &\ldots &\partial_{\mu_1}\Phi_N\\
    \vdots               &       &\vdots\\
    \partial_{\mu_N}\Phi_1 &\ldots &\partial_{\mu_N}\Phi_N
  \end{pmatrix}
  = \epsilon^{i_1\ldots i_N}\epsilon^{b_1\ldots b_N}
  \partial_{\mu_{i_1}}\Phi_{b_1}\ldots\partial_{\mu_{i_N}}\Phi_{b_N} =0.
\end{align}
That the determinant vanishes can be proven from the relation
$\Phi^T\Phi=1$ which gives $\Phi_1=\sqrt{1-\sum_{i=2,N}\Phi_i^2}$. With that  the determinant in \eqref{eq:determinant} becomes (pulling out an overall factor)
\begin{align}
  \frac{-1}{\sqrt{1-\sum_{i=2,N}\Phi_i^2}}
    \det\begin{pmatrix}\sum_{i=2,N}\Phi_i\partial_{\nu_1}\Phi_i &\partial_{\nu_1}\Phi_2 &\ldots &\partial_{\nu_1}\Phi_N\\
    \vdots &\vdots               &       &\vdots\\
    \sum_{i=2,N}\Phi_i\partial_{\nu_N}\Phi_i & \partial_{\nu_N}\Phi_2 &\ldots &\partial_{\nu_N}\Phi_N
\end{pmatrix}.
\end{align}
The first column is a linear combination of the remaining columns so
the determinant vanishes. Note that the determinant vanishes independent of the dimension, it does not require $D=N$. The Levi-Civita tensor refers to the numbering of the indices in the set $\{\mu_i\}$.

In fact, at higher orders of derivatives there are more consequences of the identity in (\ref{eq:determinant}). One can multiply it with anything that contains the same set of Lorentz indices $\{\mu_i\}$ and possibly extra derivatives, as well as many more fields. In particular the square of the determinant affects type 1 terms.

\subsection{Some comments on the implementation}

The implementation generates all possible terms using \textsc{FORM}~\cite{Vermaseren:2000nd}, \textsc{Python} and a \textsc{C++} implementation of Gaussian reduction using \texttt{gmp} \cite{Granlund:2016}.

The programs are written separately for each type but follow the same pattern. 

First we generate all possible terms of a given type distributing derivatives in all possible ways over the factors of $\Phi$ present. For each type and number of derivatives there is a maximum number of $\Phi$ that need to be taken into account.
At this stage we also use $\Phi^T\Phi=1$ and $\Phi^T\partial_\mu\Phi=f_{~;\mu}=0$. This leaves us with a large number of terms, of which many are still equivalent.
As an example, take
\begin{align}
  \Phi^T\Phi_{\mu\nu\rho}\Phi^T_\mu\Phi_{\nu\rho},~
  \Phi^T\Phi_{\mu\nu\rho}\Phi^T_\nu\Phi_{\mu\rho}
  \text{~and~}
  \Phi^T\Phi_{\mu\nu\rho}\Phi^T_\rho\Phi_{\mu\nu}
\end{align}
these are all the same term but letting a program automatically identify this does not always work (i.e. \textsc{FORM}'s command \texttt{renumber 1} together with nested symmetric functions failed to see that these were equal.)
This redundancy is removed by explicitly generating all permutations of Lorentz indices from terms to check which others are generated and removing them. By adding all permutations of Lorentz indices there will also be zeros occurring. These are also kept track of at this stage. 

The second stage corresponds to generating all possible remaining relations. We generate all terms containing two or more derivatives acting on $\Phi^T\Phi=1$, all possible total derivatives (thus taking care of partial derivation), all terms containing the lowest order equation of motion (thus taking care of possible field redefinitions\footnote{The exception is the $D=2$, $N=3$ case where the equation of motion not including the topological term has been used.}), all terms containing the determinant (\ref{eq:determinant}) and all possible antisymmetric combinations of $D+1$ Lorentz-indices and $N+1$ flavor indices.

The third stage is then to take all these relations and remove redundant ones. This is done by Gaussian elimination with a sparse matrix implementation in \textsc{C++} and \texttt{gmp}. The cases run are limited by the memory and running time (we have only run the cases taking less than a day on a PC). Choosing which terms to eliminate is not unique. We have preferentially eliminated terms with fewer factors of $\Phi$ and occurrences of $\partial^2$.

\subsection{Type 1 terms}

These terms are even under parity and intrinsic parity. A number of general considerations can be made before generating all possible terms and relations.
\begin{itemize}
\item Terms with only one $\Phi^T_{\mu_1\ldots}\Phi_{\nu_1\ldots}$ can always have, by partial integration, all derivatives moved to the same $\Phi$ and thus can be removed using field redefinitions.
\item Since the Lorentz indices are contracted in pairs, $n_d$ must be even and we have $n_d/2$ Lorentz indices.
\item Since each pair $\Phi^T_{\mu_1\dots}\Phi_{\nu_1\dots}$ must contain at least two derivatives, there are at most $n_d/2$ pairs of the form $f_{\mu_1\ldots;\nu_2\ldots}$ and we have thus at most $n_d/2$ flavor indices.
\item The determinant condition only applies if there are at least $N$ different flavor indices requiring $n_d/2\ge N$. 
There must also be at least $N$ different values for the derivatives requiring $n_d/2\ge N$ and $D\ge N$.
\item  As soon as each contracted Lorentz index pair can have a different spacetime direction, no constraints from dimensionality of spacetime apply. This occurs for $D\ge n_d/2$ or equivalently the Schouten constraints in Lorentz indices apply if $n_d/2\ge D+1$.
\item
The flavor Schouten constraints require $n_d/2 \ge N+1$.
\end{itemize}

The results for the number of terms are shown in Tab.~\ref{tab:resultstype1}. The requirements for the last three constraints to appear are clearly visible in the pattern in the table.

\begin{table}[!ht]
\centering
  \begin{tabular}[t]{rrrr}
    $n_d$ & $D$ & $N$   & \#terms\\
    \hline\hline
    2 & $\ge$2 & $\ge$2 &  1\\
    4 & $\ge$2 &      2 &  1\\
      &        & $\ge$3 &  2\\
    6 &      2 &      2 &  1\\
      &        & $\ge$3 &  3\\
      & $\ge$3 &      2 &  2\\
      &        &      3 &  4\\
      &        & $\ge$4 &  5\\
    8 &      2 &      2 &  3\\
      &        &      3 &  8\\
      &        & $\ge$4 &  9\\
      &      3 &      2 &  4\\
      &        &      3 & 12\\
      &        & $\ge$4 & 15\\
      & $\ge$4 &      2 &  4\\
      &        &      3 & 13\\
      &        &      4 & 16\\
      &        & $\ge$5 & 17\\
        \hline
  \end{tabular}
  \begin{tabular}[t]{rrrr}
    $n_d$ & $D$ & $N$ & \#terms\\
   \hline \hline
    10 &      2&      2 & 3 \\
      &        &      3 & 14\\ 
      &        & $\ge$4 & 16\\
      &      3 &      2 &  7\\
      &        &      3 & 34\\
      &        &      4 & 48\\
      &        & $\ge$5 & 49\\
      &      4 &      2 &  7\\
      &        &      3 & 38\\
      &        &      4 & 55\\
      &        & $\ge$5 & 58\\
      & $\ge$5 &      2 &  8\\
      &        &      3 & 39\\
      &        &      4 & 57\\
      &        &      5 & 60\\
    &        & $\ge$6 & 61\\
    \hline
  \end{tabular}
  \begin{tabular}[t]{rrrr}
    $n_d$ & $D$ & $N$ & \#terms\\
   \hline \hline
  12 &     2 &      2 &   7 \\
     &       &      3 &  34 \\
     &       &      4 &  45 \\
     &       & $\ge$5 &  46 \\
     &     3 &      2 &  17 \\
     &       &      3 & 114 \\
     &       &      4 & 185 \\
     &       & $\ge$5 & 193 \\
   &     4 &      2 &  20\\ 
   &       &      3 & 147\\ 
   &       &      4 & 253\\ 
   &       &      5 & 275\\ 
   &       & $\ge$6 & 276\\ 
   &     5 &      2 &  21\\ 
   &       &      3 & 153\\ 
   &       &      4 & 264\\ 
   &       &      5 & 289\\ 
   &       & $\ge$6 & 292\\ 
   & $\ge$6 &      2 &  21\\
   &        &      3 & 154\\
   &        &      4 & 265\\
   &        &      5 & 291\\
   &        &      6 & 294\\
   &        & $\ge$7 & 295\\
\hline
  \end{tabular}
  \caption{Results for type 1 terms up to $n_d=12$ obtained both with the Hilbert series method and with the explicit construction method. For an empty element, take the value above it in the same column. $n_d$ must be even.}
\label{tab:resultstype1}
\end{table}

We can also look at the Lagrangian terms that show up. At the four derivative level we started with 10 terms after stage 1 and ended up in general with the two possible terms\footnote{Known since a long time, see e.g. \cite{Gasser:1983yg}.}.
For $N=2$ the two terms are related and one can drop either of them.
\begin{table}[!ht]
    \centering
    \begin{tabular}{rll}
    $n_d$ & operator & present\\
    \hline\hline
    2 & $f_{\mu;\mu}$ & all\\
    4 & $f_{\mu;\mu} f_{\nu;\nu}$ & all\\
      &  $f_{\mu;\nu} f_{\mu;\nu}$ & $N\ge3$\\
   6 & $f_{\mu;\nu\rho}f_{\nu;\mu\rho}$ & $D\ge3$ and $N\ge3$\\
     & $f_{\mu;\nu}f_{\mu\rho;\nu\rho}$ & $D\ge3$ or $N\ge3$\\
     & $f_{\mu;\mu}f_{\nu;\nu}f_{\rho;\rho}$ & $D\ge3$ and $N\ge4$\\
     & $f_{\mu;\mu}f_{\nu;\rho}f_{\nu;\rho}$ & $N\ge3$\\
     & $f_{\mu;\nu}f_{\nu;\rho}f_{\rho;\mu}$ & all\\
 8 & $f_{\mu\nu;\mu\rho}f_{\sigma\nu;\sigma\rho}$ & $D\ge3$ and $N\ge3$\\
   & $f_{\mu\nu;\rho\sigma}f_{\mu\nu;\rho\sigma}$ & $N\ge3$ \\
   & $f_{\mu\nu;\rho\sigma}f_{\mu\rho;\nu\sigma}$ & all\\
   & $f_{\mu;\mu} f_{\nu;\rho\sigma} f_{\sigma;\nu\rho}$ & $D\ge4$ and $N\ge4$\\
   & $f_{\mu;\nu}f_{\rho;\mu\rho}f_{\sigma;\nu\sigma}$ & $D\ge3$ and $N\ge4$\\
   & $f_{\mu;\nu}f_{\rho;\mu\sigma}f_{\rho;\nu\sigma}$ & ($D\ge3$ and $N\ge4$) or ($D\ge4$ and $N\ge3$)\\
   & $f_{\mu;\nu}f_{\rho;\mu\sigma}f_{\sigma;\nu\rho}$ & $N\ge3$\\
   & $f_{\mu;\nu}f_{\rho;\rho}f_{\mu\sigma;\nu\sigma}$ & $D\ge3$ and $N\ge3$\\
   & $f_{\mu;\nu}f_{\mu;\nu}f_{\rho\sigma;\rho\sigma}$ & $D\ge3$ and $N\ge3$\\
   & $f_{\mu;\nu}f_{\mu;\rho}f_{\nu\sigma;\rho\sigma}$ & ($D\ge2$ and $N\ge4$) or $D\ge3$\\
   & $f_{\mu;\nu}f_{\rho;\sigma}f_{\mu\nu;\rho\sigma}$ & $N\ge3$\\
   & $f_{\mu;\nu}f_{\rho;\sigma}f_{\mu\rho;\nu\sigma}$ & all \\
   & $f_{\mu;\mu}f_{\nu;\nu}f_{\rho;\rho}f_{\sigma;\sigma}$ & $D\ge4$ and $N\ge5$\\
   & $f_{\mu;\mu}f_{\nu;\nu}f_{\rho;\sigma}f_{\rho;\sigma}$ & $D\ge3$ and $N\ge4$\\
   & $f_{\mu;\mu}f_{\nu;\rho}f_{\rho;\sigma}f_{\sigma;\nu}$ & $N\ge3$\\
   & $f_{\mu;\nu}f_{\mu;\nu}f_{\rho;\sigma}f_{\rho;\sigma}$ & $N\ge3$\\
   & $f_{\mu;\nu}f_{\nu;\rho}f_{\rho;\sigma}f_{\sigma;\mu}$ & all\\
\hline
    \end{tabular}
    \caption{The operators of type 1 that form a minimal Lagrangian and when they appear. "and" means both conditions need to be satisfied; "or" that one of them is sufficient. Brackets have their usual logical meaning.}
    \label{tab:operatorstype1}
\end{table}

To give an indication of the difficulty we give here the starting number of terms and the final number.
At the six derivative level we start with 48 terms after stage 1 and end up with 5 terms in the general case.
At the eight derivative level we start with 279 terms after stage 1 and end up with 17 terms in the general case.
At the ten derivative level we start with 1774 terms after stage 1 and end up with 61 terms in the general case.
At the twelve derivative level we start with 12872 terms after stage 1 and end up with 295 terms in the general case.

The operators and for which cases they appear up to $n_d=8$ are shown in Tab.~\ref{tab:operatorstype1} and up to $n_d=12$ in the supplementary file \texttt{Lagrangiansfulltype1.txt}.

\subsection{Type 2 terms}

These are terms where the group invariants are all of the type $\Phi^T_{\nu_1\dots}\Phi^{\mu_1\ldots}$, there is one factor of $\epsilon_{\mu_1\ldots\mu_D}$ with associated derivatives and there are possibly extra pairs of derivatives with contracted  Lorentz indices.

A number of constraints can be immediately derived:
\begin{itemize}
\item Terms with only one $\Phi^T_{\mu_1\ldots}\Phi_{\nu_1\ldots}$ can always have by partial integration all derivatives moved to the same $\Phi$ and thus vanish since partial derivatives commute.
\item For the Levi-Civita tensor to be fully used requires $n_d\ge D$ to have at least $D$ different Lorentz indices.
\item Lorentz invariance requires that $n_d-D$ is even since the derivatives not connected to the Levi-Civita tensor must be contracted in pairs. There are $D+(n_d-D)/2$ different Lorentz indices.
\item The derivatives corresponding to $\epsilon_{\mu_1\ldots\mu_D}$ must all act on different $\Phi$ since partial derivatives commute.
There are thus at least $D$ $\Phi$ around.
There must be extra derivatives around since $\Phi^T_{\mu_1}\Phi_{\mu_2}$ is symmetric in $\mu_1\leftrightarrow\mu_2$ and $\Phi^T\Phi_{\mu_1}=0.$
This requires $n_d\ge D+2$ for $D\le4$, $n_d\ge D+4$ for $5\le D\le8$, $n_d\ge D+6$ for $9\le D\le 12$,\ldots 
\item Since each pair $\Phi^T_{\mu_1\dots}\Phi_{\nu_1\dots}$ must contain at least two derivatives, there are at most $n_d/2$ pairs of the form $f_{\mu_1\ldots;\nu_2\ldots}$ and we have thus at most $n_d/2$ flavor indices. The actual maximum number is lower because of the argument in the previous item.
\item The Schouten identity in Lorentz indices needs at least $D+1$ different Lorentz indices or applies if $n_d-2\ge D$.
\item The Schouten identity in flavor indices requires $N+1$ different flavor indices and thus applies for $n_d/2\ge N+1$.
\end{itemize}
The results for the total number of terms are given in Tab.~\ref{tab:resultstype2}. Note that the patterns when constraints apply and the zeros following from the above arguments are clearly visible.

A choice of operators that form a  minimal basis is given in Tab.~\ref{tab:operatorstype2} for $n_d\le8$ and up to $n_d=10$ in the supplementary file \texttt{Lagrangiansfulltype2.txt}.
\begin{table}[!ht]
\centering
  \begin{tabular}[t]{rrrr}
    $n_d$ & $D$ & $N$   & \#terms \\
    \hline\hline
    2 & all & all & 0 \\
    3 & all & all & 0 \\
    4 & all & all & 0 \\
    5 & 3 &     2 & 0 \\
      &   & $\ge$3& 1 \\
      & $\ge$5 & all & 0\\
    6 & all & all & 0\\ 
    7 & 3 & 2 & 0\\
      &   & $\ge$3 & 3\\
      & $\ge$5 & all & 0\\
    8 & 2 & 2 & 1\\
      &   & $\ge$3 & 2\\
      & 4 & 2 & 0\\
      &   & 3 & 1\\
      &   & $\ge$4 & 2\\
    & $\ge$6 & all & 0\\
    9 & 3 & 2 & 2\\
      &   & 3 & 11\\
      &   &$\ge$4 & 15\\
    & $\ge$5 & all & 0\\
            \hline
  \end{tabular}
  \begin{tabular}[t]{rrrr}
    $n_d$ & $D$ & $N$   & \#terms \\
    \hline\hline
   10 &  2 & 2 &1\\
      &    & $\ge3$& 5\\
      & 4 & 2 & 0\\
      &   & 3 & 7\\
      &   & 4 & 14\\
      &   &$\ge$5 & 15\\
   &$\ge$6& all & 0\\
   11 & 3 & 2 & 3\\
       &  & 3 & 42\\
       &  & 4 & 63\\
       &  & $\geq5$ & 64\\
       & 5 & $\leq3$ & 0\\
       &  & 4 & 7\\
       &  & $\geq5$ & 8\\
       & 7 & $\leq4$ & 0\\
       &  & $\geq5$ & 1\\
       & $\geq9$ & all & 0\\
          \hline
  \end{tabular}
  \begin{tabular}[t]{rrrr}
    $n_d$ & $D$ & $N$   & \#terms \\
    \hline\hline
    12 & 2 & 2 & 3\\
       &  & 3 & 17\\
       &  & $\geq4$ & 22\\
       & 4 & 2 & 1\\
       &  & 3 & 48\\
       &  & 4 & 115\\
       &  & $\geq5$ & 125\\
       & 6 & $\leq3$ & 0\\
       &  & 4 & 5\\
       &  & $\geq5$ & 10\\
       & $\geq8$ & all & 0\\
        \hline
  \end{tabular}
  \caption{Results for type 2 terms, obtained with both the explicit construction method and the Hilbert series method up to $n_d=10$ and only with the Hilbert series method for $n_d=11,12$. For an empty column, take the value above it. Many of the zeros are explained by the arguments in the text. Note that $n_d-D$ must be even.}
\label{tab:resultstype2}
\end{table}

\begin{table}[!ht]
    \centering
    \begin{tabular}{rrll}
    $n_d$ & $D$ & operator & present\\
    \hline\hline
    5 & 3 & $\epsilon_{\mu\rho\sigma}f_{\mu;\nu}f_{\rho;\nu\sigma}$ & $N\ge3$\\
    7 & 3 & $\epsilon_{\mu\nu\tau}f_{\mu;\nu\rho}f_{\rho\sigma;\sigma\tau}$ & $N\ge3$\\
      &  & $\epsilon_{\mu\sigma\tau}f_{\mu;\nu}f_{\nu;\rho}f_{\sigma;\rho\tau}$ & $N\ge3$\\
      &  & $\epsilon_{\mu\rho\tau}f_{\mu;\nu}f_{\rho;\sigma}f_{\nu;\sigma\tau}$ & $N\ge3$\\
    8 & 2 & $\epsilon_{\mu\sigma}f_{\mu;\nu}f_{\rho;\nu\sigma}f_{\tau;\rho\tau}$ & $N\ge2$\\
      &   & $\epsilon_{\mu\tau}f_{\mu;\nu}f_{\rho;\sigma}f_{\nu\rho;\sigma\tau}$ & $N\ge3$\\
      & 4 & $\epsilon_{\mu\rho\tau\alpha}f_{\mu;\nu}f_{\rho;\nu\sigma}f_{\tau;\tau\alpha}$ & $N\ge4$\\
      &  & $\epsilon_{\mu\rho\tau\alpha}f_{\mu;\nu}f_{\rho;\sigma}f_{\nu\tau;\sigma\alpha}$ & $N\ge3$\\
\hline
    \end{tabular}
    \caption{The operators of type 2 that form the minimal Lagrangians and when they appear.}
    \label{tab:operatorstype2}
\end{table}

\subsection{Type 3 terms}

These are the terms containing one $\epsilon^{a_1\ldots a_N}$. There are a number of immediate observations:
\begin{itemize}
 \item Terms with at most two $\Phi_{a;\mu_1\ldots}$ can always have by partial integration all derivatives moved to the same $\Phi$ and thus can be removed using field redefinitions.
\item Since the Lorentz indices are contracted in pairs, $n_d$ must be even and we have $n_d/2$ Lorentz indices.   
\item The $N$ $\Phi_{a;\mu_1\ldots}$ contracted with  $\epsilon^{a_1\ldots a_N}$ must all have different combinations of derivatives operating on it. It can have no derivatives, one from each contracted indices and then the remaining ones can be pairs from the remaining derivatives. This requires $1+(n_d/2)+[n_d/4]\ge N$. The same argument but using different values of the spacetime index requires $1+D+D(D+1)/2\ge N$, but for large values of $n_d$ also three and more derivatives become allowed so the constraint on $D$ is not universally valid.
\item Extra pairs of $\Phi^T_{\mu_1\ldots}\Phi_{\nu_1\ldots}$ can be present if the number of derivatives is sufficiently much larger than $N$. The parts contracted with $\epsilon^{a_1\ldots a_N}$ require at least $N-1$ derivatives so there are at most $[(n_d-N+1)/2]$ extra $\Phi^T \Phi$ pairs.
There are thus at most $N+[(n_d-N+1)/2]$ flavor indices and at least $N$.
\item The determinant condition requires at least $N$ flavor indices and $D\ge N$ and $n_d/2\ge N$ to be present.
\item The Schouten identity in Lorentz indices requires $n_d/2\ge D+1$ to be present.
\item The Schouten identity in flavor requires at least $N+1$ different flavor indices and thus implies $[(n_d-N+1)/2]\ge 1$. 
\end{itemize}

Results for the number of terms are given in Tab.~\ref{tab:resultstype3} and the minimal Lagrangians in Tab.~\ref{tab:operatorstype3} up to $n_d=8$. The terms up to $n_d=10$ are given in the supplementary file \texttt{Lagrangiansfulltype3.txt}.

\begin{table}[!t]
    \centering
    \begin{tabular}[t]{rrrr}
    $n_d$ & $D$ & $N$ & \#terms\\
    \hline\hline
    2 & all & all & 0 \\
    4 & all & all & 0 \\
    6 & 2 & 2 &0   \\
      &   & 3 &0  \\
      &   & 4 & 1  \\
      &   & 5 & 1  \\
      &   & $\ge$6 & 0\\
      & $\ge$3 & 2 & 0\\
      &        & 3 & 0\\
      &        & 4 & 1\\
      &        & 5 & 1\\
      &    &$\ge$6& 0 \\
    8 & 2 & 2 & 0     \\
      &   & 3 & 1     \\
      &   & 4 & 2     \\
      &   & 5 & 1     \\
      &   & $\ge$6 & 0\\
      & 3 & 2 & 0     \\
      &   & 3 & 2\\
      &   & 4 & 4\\
      &   & 5 & 2\\
      & & $\ge$6 & 0\\
      & $\ge$4 & 2 & 1\\
      &        & 3 & 2\\
      &        & 4 & 4\\
      &        & 5 & 2\\
    &        &$\ge$6 & 0\\
      \hline
    \end{tabular}
    \begin{tabular}[t]{rrrr}
    $n_d$ & $D$ & $N$ & \#terms\\
    \hline\hline
   10 & 2 & 2 & 0\\  
      &   & 3 & 4\\         
      &   & 4 & 9\\         
      &   & 5 & 5\\         
      & &$\ge$6&0\\       
      & 3 & 2 & 2\\    
      &   & 3 & 14\\   
      &   & 4 & 24\\   
      &   & 5 & 14\\   
      &   & 6 & 3\\    
      &   &$\ge$7 & 0\\   
  &$\ge$4& 2 & 3\\   
  &      & 3 &16\\   
  &      & 4 & 27\\  
  &      & 5 & 17\\  
  &      & 6 & 4\\   
  &      &$\ge$7& 0\\
  12  & 2 & 2 & 3\\
 &  & 3 & 15\\
 &  & 4 & 21\\
 &  & 5 & 10\\
 &  & 6 & 1\\
 &  & $\geq7$ & 0\\
 & 3 & 2 & 8\\
 &  & 3 & 69\\
 &  & 4 & 121\\
 &  & 5 & 79\\
 &  & 6 & 22\\
 &  & 7 & 3\\
 &  & 8 & 1\\
 &  & $\geq9$ & 0\\
      \hline
    \end{tabular}
    \begin{tabular}[t]{rrrr}
    $n_d$ & $D$ & $N$ & \#terms\\
    \hline\hline
12 & 4 & 2 & 12\\
 &  & 3 & 91\\
 &  & 4 & 160\\
 &  & 5 & 116\\
 &  & 6 & 40\\
 &  & 7 & 7\\
 &  & 8 & 2\\
 &  & 9 & 1\\
 &  & $\geq10$ & 0\\
 & 5 & 2 & 12\\
 &  & 3 & 94\\
 &  & 4 & 165\\
 &  & 5 & 120\\
 &  & 6 & 41\\
 &  & 7 & 7\\
 &  & 8 & 2\\
 &  & 9 & 1\\
 &  & $\geq10$ & 0\\
 & $\geq6$ & 2 & 13\\
 &  & 3 & 94\\
 &  & 4 & 165\\
 &  & 5 & 120\\
 &  & 6 & 41\\
 &  & 7 & 7\\
 &  & 8 & 2\\
 &  & 9 & 1\\
 &  & $\geq10$ & 0\\
    \hline
    \end{tabular}
    \caption{Results for type 3 terms, obtained with both the explicit construction method and the Hilbert series method up to $n_d=10$ and only with the Hilbert series method for $n_d=12$. For an empty column take the value above it. $n_d$ must be even.}
    \label{tab:resultstype3}
\end{table}

\begin{table}[!ht]
    \centering
    \begin{tabular}{rrll}
    $n_d$ & $N$ & operators & present\\
    \hline\hline
    6 & 4 & $g_{~;\mu;\nu;\mu\rho}f_{\nu;\rho}$ & $D\ge2$\\
      & 5 & $g_{~;\mu;\nu;\mu\rho;\nu\rho}$ & $D\ge2$\\
    8 & 2 & $g_{~;\mu\nu;}f_{\rho;\sigma}f_{\mu\rho;\nu\sigma}$ & $D\ge4$\\
      & 3 & $g_{~;\mu;\nu\rho}f_{\nu;\sigma}f_{\mu;\rho\sigma}$ &$D\ge3$\\
      &   & $g_{~;\mu;\nu\rho}f_{\nu;\sigma}f_{\rho;\mu\sigma}$ & $D\ge2$\\
      & 4 & $g_{~;\mu;\nu;\mu\rho}f_{\nu\sigma;\rho\sigma}$ & $D\ge3$\\
      &  & $g_{~;\mu;\nu\rho;\nu\sigma}f_{\rho;\mu\sigma}$ & $D\ge2$\\
      &  & $g_{~;\mu;\nu;\mu\rho}f_{\rho;\sigma}f_{\nu;\sigma}$ & $D\ge2$\\
      &  & $g_{~;\mu;\nu;\rho}f_{\mu;\sigma}f_{\nu;\rho\sigma}$ & $D\ge3$\\
      & 5 & $g_{~;\mu;\nu;\mu\rho;\rho\sigma}f_{\nu;\sigma}$ & $D\ge2$\\
      &   & $g_{~;\mu;\nu;\rho;\mu\sigma}f_{\nu;\rho\sigma}$ & $D\ge3$\\
      \hline
    \end{tabular}
    \caption{The operators of type 3 that form a minimal Lagrangian and when they appear.}
    \label{tab:operatorstype3}
\end{table}

\subsection{Type 4 terms}

These are the terms containing one $\epsilon_{\mu_1\ldots\mu_D}$ with associated derivatives and one $\epsilon^{a_1\ldots a_N}$ with associated factors of $\Phi_{a_i;\mu_1\ldots}$. They are odd under parity and intrinsic parity.

There are a number of immediate observations:
\begin{itemize}
    \item $n_d-D$ must be even since the Lorentz indices not connected to $\epsilon_{\mu_1\ldots\mu_D}$ are contracted in pairs. The number of different Lorentz indices is $D+(n_d-D)/2=(n_d+D)/2$
    \item $n_d\ge D$ since $\epsilon_{\mu_1\ldots\mu_D}$ must all have different derivatives.
    \item The $\epsilon^{a_1\ldots a_N}$ flavor indices must all be connected to different $\Phi_{a;\mu\ldots}$. Using the same argument as for type 3 terms this leads to $1+(n_d+D)/2+[(n_d+D)/4]\ge N$, again not valid for large $n_d$ when also three derivatives can act on the same $\Phi$. It also requires $n_d\ge N-1$.
    \item At least $N-1$ derivatives are needed for the $\Phi_a$ connected to $\epsilon^{a_1\ldots a_N}$, so there are at most $[(n_d-N+1)/2]$ extra pairs of $\Phi^T_{\mu\ldots}\Phi_{\nu\ldots}$ present and hence the maximum number of flavor indices is $N+[(n_d-N+1)/2]$.
    \item The determinant constraint is present for $D\ge N$.
    \item The Schouten identity for Lorentz indices requires $D+1$ different Lorentz indices, so it is present for $n_d\ge D+2$. 
    \item The Schouten identity for flavor indices requires $N+1$ different flavor indices, so it is present for $[(n_d-N+1)/2]\ge1$.
    \end{itemize}

The results for the number of terms are shown in
Tabs.~\ref{tab:resultstype4} and \ref{tab:resultstype4b}.

\begin{table}[!t]
    \centering
    \begin{tabular}[t]{rrrr}
    $n_d$ & $D$ & $N$ &\#terms\\
    \hline\hline
    2 & 2 & 2 & 0\\
      &   & 3 & 1\\
      &   &$\ge$4& 0\\
     &$\ge$3 &$\ge$2& 0\\
     3 & 3 & $\le$3 & 0\\
      &   & 4 & 1\\
      &   &$\ge$5& 0\\
      &$\ge$4 &$\ge$2& 0\\
    4 & 2 & 2 & 0\\
      &   & 3 & 1\\
      &   &$\ge$4& 0\\
      & 4 &$\le$4 & 0\\
      &   & 5 & 1\\
      &   &$\ge$6&0\\
      &$\ge5$& $\ge$2 & 0\\
    5 & 3 & $\le$3&0\\
      &   & 4 &1\\
      &   & $\ge$5 &0\\
      & 5 & $\le$5 & 0\\
      &   & 6 & 1\\
      &   & $\ge$7 & 0\\
      & $\ge$7 & $\ge$2 & 0\\
   6  & 2 & 2 & 0\\
      &   & 3 & 2\\
      &   & 4 & 1\\
      &   & $\ge$5 & 0\\
      & 4 & $\le$4 & 0\\
      &   & 5 & 1\\
      &   & $\ge$6 & 0\\
      & 6 & $\le$6 & 0\\
      &   & 7 & 1\\
      &   & $\ge$8 & 0\\
      & $\ge$8 & $\ge$2 & 0\\
      \hline
    \end{tabular}
    \begin{tabular}[t]{rrrr}
    $n_d$ & $D$ & $N$ &\#terms\\
    \hline\hline
    7 & 3 &$\le3$ & 0\\
      &   & 4 & 4\\
      &   & 5 & 2\\
      &   &$\ge$6 & 0\\
      & 5 & $\le$5 & 0\\
      &  & 6 & 1\\
      &  &$\ge$7 & 0\\
      &7 &$\le$7 & 0\\
      &  & 8 & 1\\
      &  &$\ge$9 & 0\\
      &$\ge$9 & $\ge$2 & 0\\
    8 & 2 & 2 & 0\\
      &  & 3 & 5\\
      &  & 4 & 2\\
      &  &$\ge$5& 0\\
      & 4 & $\le$4 & 0\\
      &   & 5 & 4\\
      &   & 6 & 2\\
      &   & $\ge$7 & 0\\
      & 6 & $\le$6 & 0\\
      &   & 7 & 1\\
      &   &$\ge$9 & 0\\
      & 8 & $\le$8 & 0\\
      &   & 9 & 1\\
      &   & $\ge$10 & 0\\
      & $\ge$10& $\ge$2 & 0\\
      \hline
      \end{tabular}
      \begin{tabular}[t]{rrrr}
    $n_d$ & $D$ & $N$ &\#terms\\
    \hline\hline
    9 & 3 & 2 &0\\
      &   & 3 & 4\\
      &   & 4 & 15\\
      &   & 5 & 8\\
      &   &$\ge$6 & 0\\
      & 5 & 2 & 0\\
      &   & 3 & 1\\
      &   & 4 & 0\\
      &   & 5 &0\\
      &   & 6 & 4\\
      &   & 7 & 2\\
      &   &$\ge$8 & 0\\
      & 7 & $\le$7 & 0\\
      &   & 8 & 1\\
      &   &$\ge$9 & 0\\
      & 9 & $\le$9 & 0\\
      &   & 10 & 1\\
      &   &$\ge$11 & 0\\
      & $\ge$11 & $\ge$2 & 0\\
      \hline
      \end{tabular}
\caption{Results for the number of terms of type 4, obtained with both the explicit construction method and the Hilbert series method. For an empty column take the value above it. $n_d-D$ must be even. }
    \label{tab:resultstype4}
\end{table}

\begin{table}
    \centering
      \begin{tabular}[t]{rrrr}
    $n_d$ & $D$ & $N$ &\#terms\\
    \hline\hline
    10 & 2 & 2 &0\\
      &   & 3 & 11\\
      &   & 4 & 9\\
      &   & 5 & 1\\
      &   &$\ge$6&0\\
      & 4 & 2 & 1\\
      &   & 3 & 3\\
      &   & 4 & 11\\
      &   & 5 & 20\\
      &   & 6 & 9\\
      &   & $\ge$7 & 0\\
      & 6 & $\le$3& 0\\
      &   & 4 & 1\\
      &   & 5 & 0\\
      &   & 6 & 0\\
      &   & 7 & 4\\
      &   & 8 & 2\\
      &   &$\ge$9& 0\\
      & 8 & $\ge$8 & 0\\
      &   & 9 & 1\\
      &   & $\le$10&0\\
      & 10& $\le$10& 0\\
      &   & 11 & 1\\
      &   &$\ge$11&0\\
      &$\ge12$& all &0\\
      \hline
      \end{tabular}
      \begin{tabular}[t]{rrrr}
    $n_d$ & $D$ & $N$ &\#terms\\
    \hline\hline
    11 & 3 & 2 & 0\\
 &  & 3 & 25\\
 &  & 4 & 65\\
 &  & 5 & 39\\
 &  & 6 & 5\\
 &  & $\geq7$ & 0\\
 & 5 & 2 & 0\\
 &  & 3 & 5\\
 &  & 4 & 7\\
 &  & 5 & 13\\
 &  & 6 & 22\\
 &  & 7 & 10\\
 &  & $\geq8$ & 0\\
 & 7 & $\leq4$ & 0\\
 &  & 5 & 1\\
 &  & 6 & 0\\
 &  & 7 & 0\\
 &  & 8 & 4\\
 &  & 9 & 2\\
 &  & $\geq10$ & 0\\
 & 9 & $\leq9$ & 0\\
 &  & 10 & 1\\
 &  & $\geq11$ & 0\\
 & 11 & $\leq11$ & 0\\
 &   & 12 & 1\\
 &   & $\geq13$ & 0\\
 & $\geq13$ & all & 0\\
    \hline
    \end{tabular}
    \begin{tabular}[t]{rrrr}
    $n_d$ & $D$ & $N$ &\#terms\\
    \hline\hline
    12 & 2 & 2 & 3\\
 &  & 3 & 27\\
 &  & 4 & 21\\
 &  & 5 & 4\\
 &  & 6 & 1\\
 &  & 7 & 1\\
 &  & $\geq8$ & 0\\
 & 4 & 2 & 2\\
 &  & 3 & 37\\
 &  & 4 & 99\\
 &  & 5 & 126\\
 &  & 6 & 66\\
 &  & 7 & 10\\
 &  & $\geq8$ & 0\\
 & 6 & 2 & 0\\
 &  & 3 & 1\\
 &  & 4 & 8\\
 &  & 5 & 8\\
 &  & 6 & 13\\
 &  & 7 & 22\\
 &  & 8 & 10\\
 &  & $\geq9$ & 0\\
 & 8 & $\leq5$ & 0\\
 &  & 6 & 1\\
 &  & 7 & 0\\
 &  & 8 & 0\\
 &  & 9 & 4\\
 &  & 10 & 2\\
 &  & $\geq11$ & 0\\
 & 10 & $\leq10$ & 0\\
 &  & 11 & 1\\
 &  & 12 & 0\\
 & 12 & $\leq12$ & 0\\
 &   & 13 & 1\\
 &   & $\geq14$ & 0\\
 & $\geq14$ & all & 0\\
    \hline
    \end{tabular}
\caption{Results for the number of terms of type 4, obtained with both the explicit construction method and the Hilbert series method for $n_d=10$ and only with the Hilbert series method for $n_d=11,12$. For an empty column take the value above it. $n_d-D$ must be even.}
    \label{tab:resultstype4b}
\end{table}

The operators of type 4 are different for each case of $D$ and $N$. They are listed in Tab.~\ref{tab:operatorstype4} up to $n_d=8$.  The operators up to $n_d=10$ are given in the supplementary file \texttt{Lagrangiansfulltype4.txt}.

For $n_d=D$ ones sees that there is only a term existing for $N=D+1$. This can be proven as follows:
For $n_d=D$ all derivatives must be coupled to $\epsilon_{\mu_1\ldots\mu_D}$. So each $\Phi_a$ can have at most one derivative acting on it because of the antisymmetry. In addition $\Phi^T_{\mu_i}\Phi_{mu_j}$ also vanishes, implying that all $\Phi_{a;\mu_i}$ must be connected to $\epsilon_{a_1\ldots a_N}$. This leaves only two possible terms $\epsilon_{\mu_1\ldots\mu_D}g_{\mu_1;\ldots;\mu_D}$ for $N=D$ and  $\epsilon_{~;\mu_1\ldots\mu_D}g_{\mu_1;\ldots;\mu_D}$ for $N=D+1$. The first case is a total derivative and as such does not contribute leaving only the second term. That this occurs is in fact a check on both the explicit construction of the Lagrangians and on the programs.

Note that there is another clear pattern visible also for $n_d=D+2$. Only for $N=D+1$ there exists a term and it is of the form $\epsilon_{\mu_1\ldots\mu_D}g_{~;\mu_1;\ldots;\mu_D}f_{\nu;\nu}$. For this we only have a partial proof. The requirement of $n_d\ge N-1$ implies that no terms exists for $N\ge D+3$. For $N=D+2$ 
only one term is not obviously zero or removable by field redefinitions, it is $\epsilon_{\mu_1\ldots\mu_D}g_{~;\mu_1;\ldots;\mu_{D-1};\mu_D\nu;\nu}$. This term is related to terms removable by a field redefinition by partial integration on the $\nu$ in $\mu_{D}\nu$. The maximum number of derivatives that can act on $\Phi$ not connected to $\epsilon_{a_1\ldots a_N}$ is of the form $\Phi^T_{\mu_1}\Phi_{\nu\mu_2}\Phi^T_{\mu_3}\Phi_{\nu\mu_4}$ allowing $N=D-4$ with a term of the form
$g_{\mu_5;\ldots;\mu_D}\Phi^T_{\mu_1}\Phi_{\nu\mu_2}\Phi^T_{\mu_3}\Phi_{\nu\mu_4}$ which is a total derivative. This proves that there are no terms for $N\le D-4$. We have no proof for the remaining 5 cases of $N$.

Note that the results of the Hilbert series also live up to these patterns.

\begin{table}[!t]
    \centering
    \begin{tabular}[t]{rrrl}
    $n_d$ & $D$ & $N$ & operators\\
    \hline\hline
    2 & 2 & 3 & $\epsilon_{\mu\nu}g_{~;\mu;\nu}$\\
    3 & 3 & 4 & $\epsilon_{\mu\nu\rho}g_{~;\mu;\nu;\rho}$\\
    4 & 2 & 3 & $\epsilon_{\mu\nu}g_{~;\mu;\nu}f_{\rho;\rho}$\\
      & 4 & 5 & $\epsilon_{\mu\nu\rho\sigma}g_{~;\mu;\nu;\rho;\sigma}$\\
    5 & 3 & 4 & $\epsilon_{\mu\nu\rho}g_{~;\mu;\nu;\rho}f_{\sigma;\sigma}$ \\
      & 5 & 6 &$\epsilon_{\mu\nu\rho\sigma\tau}g_{~;\mu;\nu;\rho;\sigma;\tau}$\\
    6 & 2 & 3 & $\epsilon_{\mu\nu}g_{~;\mu;\nu}f_{\rho;\rho}f_{\sigma;\sigma}$\\
    & & &$\epsilon_{\mu\nu}g_{~;\mu;\nu}f_{\rho;\sigma}f_{\rho;\sigma}$\\
     & & 4 & $\epsilon_{\mu\nu}g_{~;\mu;\nu;\rho\sigma}f_{\rho;\sigma}$\\
       & 4 & 5 & $\epsilon_{\mu\nu\rho\sigma}g_{~;\mu;\nu;\rho;\sigma}f_{\tau;\tau}$\\
      & 6 & 7 & $\epsilon_{\mu\nu\rho\sigma\tau\alpha}g_{~;\mu;\nu;\rho;\sigma;\tau;\alpha}$\\
     7 & 3 & 4 & $\epsilon_{\mu\sigma\tau}g_{~;\mu;\nu\rho;\nu\sigma}f_{\rho;\tau}$\\
      & & & $\epsilon_{\mu\rho\tau}g_{~;\mu;\nu\rho;\sigma\tau}f_{\nu;\sigma}$\\
      & & & $\epsilon_{\mu\nu\rho}g_{~;\mu;\nu;\rho}f_{\sigma;\sigma}f_{\tau;\tau}$\\
       & & & $\epsilon_{\mu\nu\rho}g_{~;\mu;\nu;\rho}f_{\sigma;\tau}f_{\sigma;\tau}$\\
        & & 5 & $\epsilon_{\mu\sigma\tau}g_{~;\mu;\nu\rho;\nu\sigma;\rho\tau}$\\
        & & & $\epsilon_{\mu\nu\rho}g_{~;\mu;\nu;\rho;\sigma\tau}f_{\sigma;\tau}$\\
        & 5 & 6 & $\epsilon_{\mu\nu\rho\sigma\tau}g_{~;\mu;\nu;\rho;\sigma;\tau}f_{\alpha;\alpha}$\\
        & 7 & 8 &$\epsilon_{\mu\nu\rho\sigma\tau\alpha\beta}g_{~;\mu;\nu;\rho;\sigma;\tau;\alpha;\beta}$\\
        \hline
    \end{tabular}
    \begin{tabular}[t]{rrrl}
    $n_d$ & $D$ & $N$ & operators\\
    \hline\hline
    8 & 2 & 3 & $\epsilon_{\nu\sigma}g_{~;\mu\nu;\rho\sigma}f_{\mu\tau;\rho\tau}$\\
     & & & $\epsilon_{\mu\rho}g_{~;\mu;\nu\rho}f_{\sigma;\tau}f_{\nu;\sigma\tau}$\\
     & & & $\epsilon_{\nu\sigma}g_{~;\mu\nu;\rho\sigma}f_{\mu;\tau}f_{\rho;\tau}$\\
     & & & $\epsilon_{\mu\nu}g_{~;\mu;\nu}f_{\rho;\rho}f_{\sigma;\tau}f_{\sigma;\tau}$\\
     & & & $\epsilon_{\mu\nu}g_{~;\mu;\nu}f_{\rho;\sigma}f_{\sigma;\tau}f_{\tau;\rho}$\\
     & & 4 & $\epsilon_{\mu\rho}g_{~;\mu;\nu\rho;\sigma\tau}f_{\nu;\sigma\tau}$\\
     & & & $\epsilon_{\mu\nu}g_{~;\mu;\nu;\sigma\tau}f_{\rho;\tau}f_{\sigma;\tau}$\\
     & 4 & 5 & $\epsilon_{\mu\nu\tau\alpha}g_{~;\mu;\nu;\rho\sigma;\rho\tau}f_{\sigma;\alpha}$\\
     & & & $\epsilon_{\mu\nu\sigma\alpha}g_{~;\mu;\nu;\rho\sigma;\tau\alpha}f_{\rho;\tau}$\\
     & & & $\epsilon_{\mu\nu\rho\sigma}g_{~;\mu;\nu;\rho;\sigma}f_{\tau;\tau}f_{\alpha;\alpha}$\\
      & & & $\epsilon_{\mu\nu\rho\sigma}g_{~;\mu;\nu;\rho;\sigma}f_{\tau;\alpha}f_{\tau;\alpha}$\\
    & & 6 & $\epsilon_{\mu\nu\tau\alpha}g_{~;\mu;\nu;\rho\sigma;\rho\tau;\sigma\alpha}$\\
    & & & $\epsilon_{\mu\nu\rho\sigma}g_{~;\mu;\nu;\rho;\sigma;\tau\alpha}f_{\tau;\alpha}$\\
    & 6 & 7 & $\epsilon_{\mu\nu\rho\sigma\tau\alpha}g_{~;\mu;\nu;\rho;\sigma;\tau;\alpha}f_{\beta;\beta}$\\
    & 8 & 9 & $\epsilon_{\mu\nu\rho\sigma\tau\alpha\beta\gamma}g_{~;\mu;\nu;\rho;\sigma;\tau;\alpha;\beta;\gamma}$\\
    \hline
    \end{tabular}
    \caption{The operators of type 4 for the minimal Lagrangians. Note that only cases where there is at least one term in the Lagrangian are given.}
    \label{tab:operatorstype4}
\end{table}

\section{Conclusions}\label{sec:conclusions}

In this paper, we have found the operators in a minimal basis for the $\Og(N)$ nonlinear sigma model using both the Hilbert series and explicit construction methods, for the operators with mass dimension up to 16 in spacetime dimensions up to $D=12$ and $N$ up to 12.
We find total agreement between the results of the two different methods, both in numbers of operators and forms of the operators.

We provide further evidence for the conjecture of Ref.~\cite{Henning:2017fpj}, that the exceptional Hilbert series is described completely by the co-closed but not co-exact forms -- of which there are only two in the $\Og(N)$ nonlinear sigma model, for given $D$, $N$ and mass dimension $n_d$.

In addition in the explicit construction case we proved a number of general patterns occurring in the type of operators.

It would be interesting to make similar considerations, as have been made in this paper, in the $\mathbb{C}P^{N-1}$ nonlinear sigma model or in nonlinear sigma models on flag manifolds.

\subsection*{Acknowledgments}
S.~B.~G.~thanks Guilherme Sadovski and Baiyang Zhang for discussions.
The work of J.~B.~is supported in part by the Swedish Research Council grants contract numbers 2016-05996 and 2019-03779.
S.~B.~G.~thanks the Outstanding Talent Program of Henan University and
the Ministry of Education of Henan Province for partial support.
The work of S.~B.~G.~is supported by the National Natural Science
Foundation of China (Grants No.~11675223 and No.~12071111) and by the Ministry of Science and Technology of China (Grant No.~G2022026021L).

\appendix

\section{Characters}\label{app:characters}
The dimension $k$ character of the Lorentz group \cite{Henning:2017fpj}
\beq
\sum_{k=0}^\infty 
p^k\chi_{(k,0,\cdots,0)}(x)
= \sum_{k=0}^\infty 
p^k\left(\chi_{{\rm sym}^k(\square)}(x) - p^2\chi_{{\rm sym}^k(\square)}(x)\right),
\eeq
is thus equal to the group character of the symmetric product of $k$ derivatives ($\square$ refers to the fundamental representation) with the contraction of two derivatives modded out.
This is how the lowest-order equation of motion is taken into account.
More formally, this is due to the conformal dimension of a scalar field saturating the unitarity bound and hence it furnishes a short representation of the Lorentz group.

In order to arrive at the character for the vector representation, we notice that formally, $\p_\mu\phi$ transforms as a Lorentz vector and it is sufficient to remove the scalar component of the single particle module.
This is done by summing from $k=1$ instead of $k=0$:
\begin{align}
\sum_{k=1}^\infty 
p^k\chi_{(k,0,\cdots,0)}(x)
&= \sum_{k=1}^\infty 
p^k\left(\chi_{{\rm sym}^k(\square)}(x) - p^2\chi_{{\rm sym}^k(\square)}(x)\right)\non
&=-1 + \sum_{k=0}^\infty 
p^k\left(\chi_{{\rm sym}^k(\square)}(x) - p^2\chi_{{\rm sym}^k(\square)}(x)\right)\non
&=-1 + (1-p^2)P_+(p,x),
\end{align}
which is the result of Eq.~\eqref{eq:chi_u}.
The momentum generating function $P_+(p,x)$ is defined as the group character of the symmetric product of the fundamental representation.
Considering first even dimensional space, $D=2r$, the diagonalization of an element $h\in\SO(2r)$ gives the parametrization of the maximal torus
\beq
h = \diag(x_1,x_1^{-1},\cdots,x_r,x_r^{-1}),
\label{eq:h_even}
\eeq
and the symmetric product of $k$ elements yields
\beq
\chi_{{\rm sym}^k(\square)}^{(2r)}(x) = 
\sum_{n_1+\bar{n}_1+\cdots+n_r+\bar{n}_r=k}
(x_1)^{n_1}(x_1)^{-\bar{n}_1}\cdots
(x_r)^{n_r}(x_r)^{-\bar{n}_r},\qquad
n_i,\bar{n}_i\in\mathbb{Z}_{\geq0}\;\;\forall i.
\eeq
Performing the sum over $k$ yields
\beq
P_+^{(2r)}(p,x) 
= \sum_{k=0}^\infty p^k \chi_{{\rm sym}^k(\square)}^{(2r)}(x)
=\prod_{i=1}^r\frac{1}{(1-p x_i)(1-p/x_i)}.
\eeq
For $D=2r+1$ the diagonalized element $h\in\SO(2r+1)$ has the dimension one larger than the even case, but the same number of Cartan generators (i.e.~$r$), which means the maximal torus can be parametrized as
\beq
h = \diag(1,x_1,x_1^{-1},\cdots,x_r,x_r^{-1}),
\label{eq:h_odd}
\eeq
and the symmetric product of $k$ elements yields now
\begin{equation}
\chi_{{\rm sym}^k(\square)}^{(2r+1)}(x) = 
\sum_{n_0+n_1+\bar{n}_1+\cdots+n_r+\bar{n}_r=k}
(1)^{n_0}
(x_1)^{n_1}(x_1)^{-\bar{n}_1}\cdots
(x_r)^{n_r}(x_r)^{-\bar{n}_r},\qquad
n_i,\bar{n}_i\in\mathbb{Z}_{\geq0}\;\;\forall i.
\end{equation}
Performing the sum over $k$ yields
\beq
P_+^{(2r+1)}(p,x)
=\sum_{k=0}^\infty p^k \chi_{{\rm sym}^k(\square)}^{(2r+1)}(x)
=\frac{1}{1-p}\prod_{i=1}^r\frac{1}{(1-p x_i)(1-p/x_i)}.
\eeq
Finally, we have arrived at the result in Eq.~\eqref{eq:P+}.

The group characters for $\SO(N-1)$ are simpler, since they do not involve the infinite possibilities provided by the single particle module, but are simply the group theoretic characters.
For $N-1=2r$, we have that a diagonalized group element is given by Eq.~\eqref{eq:h_even} and hence the trace is given by
\beq
\chi_{H,u}(y) = \sum_{i=1}^r(y_i + y_i^{-1}),
\eeq
where we have replaced $x_i\to y_i$ due to the group being the internal symmetry group.
Likewise for the odd case, $N-1=2r+1$, we have the element of Eq.~\eqref{eq:h_odd} and hence the trace
\beq
\chi_{H,u}(y) = 1+\sum_{i=1}^r(y_i + y_i^{-1}).
\eeq
We thus arrive at the result of Eq.~\eqref{eq:chi_Hu}.

Let us consider the case of negative (spacetime) parity in $D=2r$ dimensions.
In this case, we have to flip the last component of the vector representation. Therefore the maximal torus is parametrized by
\beq
h=\diag(x_1,x_1^{-1},\cdots,x_{r-1},x_{r-1}^{-1},1,-1).
\eeq
Clearly the first $(r-1)$ elements give rise to $P_+(p,x)$ for $D=2(r-1)$, whereas the latter two elements give $(1-p)^{-1}$ and $(1+p)^{-1}$:
\begin{align}
P_-^{2r}(p,x) 
&= \sum_{k=0}^\infty p^k \chi_{{\rm sym}^k(\square)}^{(2r),-}(x)\non
&= \sum_{k=0}^\infty p^k
\sum_{n_1+\bar{n}_1+\cdots+n_{r-1}+\bar{n}_{r-1}+n_r+\bar{n}_r=k}\!\!\!\!\!\!\!\!\!\!\!\!\!\!\!\!\!\!\!
(x_1)^{n_1}(x_1)^{-\bar{n}_1}\cdots(x_{r-1})^{n_{r-1}}(x_{r-1})^{-\bar{n}_{r-1}}(1)^{n_r}(-1)^{-\bar{n}_r}\non
&= \frac{1}{1-p}\frac{1}{1+p}P_+^{2r-2}(p,\tilde{x})\non
&= \frac{1}{1-p^2}P_+^{2r-2}(p,\tilde{x}),\qquad
\tilde{x}=(x_1,x_2,\cdots,x_{r-1}).
\end{align}
This gives the result of Eq.~\eqref{eq:P_-}.
Finally, the character function for the field $u$ with parity in $D=2r$ dimensions is obtained by taking into account an overall sign flip due to $u$ being in the vector representation:
\beq
\chi_u^{P^-}(p,x,y) =
-\left[(1-p^2)P_-^{2r}(p,x) - 1\right]\chi_{H,u}(y).
\eeq
We thus arrive at the result in Eq.~\eqref{eq:chi_u_-}.

\section{Representations}\label{app:reps}

In Eq.~\eqref{eq:w}, the field $w$ transforms under the adjoint representation of $H$, but the field $u$ belongs to the coset space $\mathfrak{g}/\mathfrak{h}$.
The adjoint representation splits under the symmetry breaking $G=\Og(N)\to H=\Og(N-1)$ as
\beq
{\rm Adj}_{N} = {\rm Adj}_{N-1} \oplus \square_{N-1},
\eeq
or in the notation of dimensions
\beq
\frac{N(N-1)}{2} = \frac{(N-1)(N-2)}{2} \oplus (N-1).
\eeq
It is thus clear that $u$ transforms according to the vector representation of $\Og(N-1)$.
An explicit matrix realization of the symmetry transformation can be written out as
\begin{align}
\begin{pmatrix}
  0 & \phi\\
  -\phi^{T} & 0
\end{pmatrix}
&\to
\begin{pmatrix}
  h & 0\\
  0 & 1
\end{pmatrix}
\begin{pmatrix}
  0 & \phi\\
  -\phi^{T} & 0
\end{pmatrix}
\begin{pmatrix}
  h^{T} & 0\\
  0 & 1
\end{pmatrix}\non
&=
\begin{pmatrix}
  0 & h\phi\\
  -\phi^{T}h^{T} & 0
\end{pmatrix}.
\end{align}
Since $\phi$ transforms according to a real representation, the two copies are identical.
For complex representations, the field $u$ would transform under the $\square\oplus\overline\square$ representation \cite{Coleman:1969sm}.

\section{Results for \texorpdfstring{$n_d>12$}{n\_d>12}}
\label{app:tables}

In this appendix, we show the results for the number of operators
of type 1 for $n_d=14,16$ in Tab.~\ref{tab:resultstype1b},
of type 2 for $n_d=13,14,15,16$ in Tabs.~\ref{tab:resultstype2b} and \ref{tab:resultstype2c}
of type 3 for $n_d=14,16$ in Tabs.~\ref{tab:resultstype3b} and \ref{tab:resultstype3c}
of type 4 for $n_d=13,14,15,16$ in Tabs.~\ref{tab:resultstype4c}, \ref{tab:resultstype4d} and \ref{tab:resultstype4e},
all obtained by the Hilbert series method.

\begin{table}
  \centering
\begin{tabular}[t]{rrrr}
    $n_d$ & $D$ & $N$ & \#terms\\
   \hline \hline
  14 & 2 & 2 & 7\\
 &  & 3 & 66\\
 &  & 4 & 94\\
 &  & $\geq5$ & 96\\
 & 3 & 2 & 34\\
 &  & 3 & 396\\
 &  & 4 & 753\\
 &  & 5 & 810\\
 &  & $\geq6$ & 812\\
 & 4 & 2 & 46\\
 &  & 3 & 602\\
 &  & 4 & 1260\\
 &  & 5 & 1431\\
 &  & $\geq6$ & 1441\\
 & 5 & 2 & 50\\
 &  & 3 & 647\\
 &  & 4 & 1377\\
 &  & 5 & 1589\\
 &  & 6 & 1613\\
 &  & $\geq7$ & 1614\\
 & 6 & 2 & 50\\
 &  & 3 & 652\\
 &  & 4 & 1387\\
 &  & 5 & 1604\\
 &  & 6 & 1631\\
 &  & $\geq7$ & 1634\\
 & $\geq7$ & 2 & 51\\
 &  & 3 & 653\\
 &  & 4 & 1389\\
 &  & 5 & 1606\\
 &  & 6 & 1634\\
 &  & 7 & 1637\\
 &  & $\geq8$ & 1638\\
   \hline
 \end{tabular}
 \begin{tabular}[t]{rrrr}
    $n_d$ & $D$ & $N$ & \#terms\\
   \hline \hline
16 & 2 & 2 & 18\\
 &  & 3 & 160\\
 &  & 4 & 258\\
 &  & $\geq5$ & 268\\
 & 3 & 2 & 85\\
 &  & 3 & 1454\\
 &  & 4 & 3285\\
 &  & 5 & 3690\\
 &  & $\geq6$ & 3706\\
 & 4 & 2 & 134\\
 &  & 3 & 2744\\
 &  & 4 & 7140\\
 &  & 5 & 8640\\
 &  & 6 & 8760\\
 &  & $\geq7$ & 8762\\
 & 5 & 2 & 149\\
 &  & 3 & 3098\\
 &  & 4 & 8325\\
 &  & 5 & 10358\\
 &  & 6 & 10611\\
 &  & $\geq7$ & 10621\\
 \hline
 \end{tabular}
 \begin{tabular}[t]{rrrr}
    $n_d$ & $D$ & $N$ & \#terms\\
   \hline \hline
 16 & 6 & 2 & 152\\
 &  & 3 & 3150\\
 &  & 4 & 8474\\
 &  & 5 & 10590\\
 &  & 6 & 10884\\
 &  & 7 & 10908\\
 &  & $\geq8$ & 10909\\
 & 7 & 2 & 153\\
 &  & 3 & 3156\\
 &  & 4 & 8486\\
 &  & 5 & 10607\\
 &  & 6 & 10906\\
 &  & 7 & 10933\\
 &  & $\geq8$ & 10936\\
 & $\geq8$ & 2 & 153\\
 &  & 3 & 3157\\
 &  & 4 & 8487\\
 &  & 5 & 10609\\
 &  & 6 & 10908\\
 &  & 7 & 10936\\
 &  & 8 & 10939\\
 &  & $\geq9$ & 10940\\
 \hline
  \end{tabular}
\caption{Results for type 1 terms with $n_d=14,16$ obtained with the Hilbert series method. For an empty element, take the value above it in the same column. $n_d$ must be even.}
  \label{tab:resultstype1b}
\end{table}

\begin{table}[!htp]
  \centering
  \begin{tabular}[t]{rrrr}
    $n_d$ & $D$ & $N$   & \#terms \\
    \hline\hline
   13 & 2 & $\geq2$ & 0\\
 & 3 & 2 & 12\\
 &  & 3 & 167\\
 &  & 4 & 299\\
 &  & $\geq5$ & 314\\
 & 4 & $\geq2$ & 0\\
 & 5 & 2 & 0\\
 &  & 3 & 21\\
 &  & 4 & 109\\
 &  & 5 & 139\\
 &  & $\geq6$ & 140\\
 & 6 & $\geq2$ & 0\\
 & 7 & 2 & 0\\
 &  & 3 & 1\\
 &  & 4 & 2\\
 &  & 5 & 9\\
 &  & $\geq6$ & 10\\
 & $\geq8$ & $\geq2$ & 0\\
 \hline
  \end{tabular}
  \begin{tabular}[t]{rrrr}
    $n_d$ & $D$ & $N$   & \#terms \\
    \hline\hline
14 & 2 & 2 & 3\\
 &  & 3 & 41\\
 &  & $\geq4$ & 58\\
 & 3 & $\geq2$ & 0\\
 & 4 & 2 & 7\\
 &  & 3 & 301\\
 &  & 4 & 806\\
 &  & 5 & 931\\
 &  & $\geq6$ & 935\\
 & 5 & $\geq2$ & 0\\
 & 6 & 2 & 0\\
 &  & 3 & 9\\
 &  & 4 & 80\\
 &  & 5 & 144\\
 &  & $\geq6$ & 151\\
 & 7 & $\geq2$ & 0\\
 & 8 & $\leq4$ & 0\\
 &  & 5 & 1\\
 &  & $\geq6$ & 3\\
 & $\geq9$ & $\geq2$ & 0\\
 \hline
  \end{tabular}
\caption{Results for type 2 terms, obtained with the Hilbert series method for $n_d=13,14$. For an empty column, take the value above it. Many of the zeros are explained by the arguments in the text. Note that $n_d-D$ must be even.}
\label{tab:resultstype2b}
    \end{table}

\begin{table}
\centering
  \begin{tabular}[t]{rrrr}
    $n_d$ & $D$ & $N$   & \#terms \\
    \hline\hline
    15 & 2 & $\geq2$ & 0\\
 & 3 & 2 & 33\\
 &  & 3 & 661\\
 &  & 4 & 1400\\
 &  & 5 & 1532\\
 &  & $\geq6$ & 1534\\
 & 4 & $\geq2$ & 0\\
 & 5 & 2 & 3\\
 &  & 3 & 255\\
 &  & 4 & 1146\\
 &  & 5 & 1566\\
 &  & $\geq6$ & 1599\\
 & 6 & $\geq2$ & 0\\
 & 7 & 2 & 0\\
 &  & 3 & 6\\
 &  & 4 & 39\\
 &  & 5 & 118\\
 &  & 6 & 141\\
 &  & $\geq7$ & 142\\
 & $\geq8$ & $\geq2$ & 0\\
 \hline
  \end{tabular}
  \begin{tabular}[t]{rrrr}
    $n_d$ & $D$ & $N$   & \#terms \\
    \hline\hline
16 & 2 & 2 & 11\\
 &  & 3 & 117\\
 &  & 4 & 191\\
 &  & $\geq5$ & 194\\
 & 3 & $\geq2$ & 0\\
 & 4 & 2 & 40\\
 &  & 3 & 1784\\
 &  & 4 & 5564\\
 &  & 5 & 6870\\
 &  & 6 & 6954\\
 &  & $\geq7$ & 6955\\
 & 5 & $\geq2$ & 0\\
 & 6 & 2 & 1\\
 &  & 3 & 155\\
 &  & 4 & 1081\\
 &  & 5 & 1955\\
 &  & 6 & 2109\\
 &  & $\geq7$ & 2112\\
 & 7 & $\geq2$ & 0\\
 & 8 & 2 & 0\\
 &  & 3 & 1\\
 &  & 4 & 9\\
 &  & 5 & 46\\
 &  & 6 & 79\\
 &  & $\geq7$ & 82\\
 & $\geq9$ & $\geq2$ & 0\\
    \hline
  \end{tabular}
\caption{Results for type 2 terms with $n_d=15,16$, obtained with the Hilbert series method. For an empty column, take the value above it. Many of the zeros are explained by the arguments in the text. Note that $n_d-D$ must be even.}
\label{tab:resultstype2c}
\end{table}

\begin{table}
    \centering
    \begin{tabular}[t]{rrrr}
    $n_d$ & $D$ & $N$ & \#terms\\
    \hline\hline
14 & 2 & 2 & 3\\
 &  & 3 & 39\\
 &  & 4 & 70\\
 &  & 5 & 38\\
 &  & 6 & 4\\
 &  & $\geq7$ & 0\\
 & 3 & 2 & 20\\
 &  & 3 & 305\\
 &  & 4 & 603\\
 &  & 5 & 439\\
 &  & 6 & 157\\
 &  & 7 & 33\\
 &  & 8 & 5\\
 &  & $\geq9$ & 0\\
 & 4 & 2 & 32\\
 &  & 3 & 478\\
 &  & 4 & 1021\\
 &  & 5 & 851\\
 &  & 6 & 369\\
 &  & 7 & 99\\
 &  & 8 & 24\\
 &  & 9 & 5\\
 &  & $\geq10$ & 0\\
\hline
    \end{tabular}
    \begin{tabular}[t]{rrrr}
    $n_d$ & $D$ & $N$ & \#terms\\
    \hline\hline
14 & 5 & 2 & 35\\
 &  & 3 & 511\\
 &  & 4 & 1098\\
 &  & 5 & 931\\
 &  & 6 & 415\\
 &  & 7 & 116\\
 &  & 8 & 29\\
 &  & 9 & 6\\
 &  & $\geq10$ & 0\\
 & $\geq6$ & 2 & 36\\
 &  & 3 & 514\\
 &  & 4 & 1102\\
 &  & 5 & 935\\
 &  & 6 & 417\\
 &  & 7 & 116\\
 &  & 8 & 29\\
 &  & 9 & 6\\
 &  & $\geq10$ & 0\\
 \hline
    \end{tabular}
    \caption{Results for type 3 terms with $n_d=14$, obtained with the Hilbert series method. For an empty column take the value above it. $n_d$ must be even.}
    \label{tab:resultstype3b}
\end{table}

\begin{table}
    \centering
    \begin{tabular}[t]{rrrr}
    $n_d$ & $D$ & $N$ & \#terms\\
    \hline\hline
16 & 2 & 2 & 10\\
 &  & 3 & 110\\
 &  & 4 & 178\\
 &  & 5 & 98\\
 &  & 6 & 19\\
 &  & 7 & 1\\
 &  & $\geq8$ & 0\\
 & 3 & 2 & 61\\
 &  & 3 & 1261\\
 &  & 4 & 2904\\
 &  & 5 & 2410\\
 &  & 6 & 1026\\
 &  & 7 & 262\\
 &  & 8 & 45\\
 &  & 9 & 4\\
 &  & $\geq10$ & 0\\
 & 4 & 2 & 109\\
 &  & 3 & 2434\\
 &  & 4 & 6379\\
 &  & 5 & 6201\\
 &  & 6 & 3254\\
 &  & 7 & 1082\\
 &  & 8 & 261\\
 &  & 9 & 48\\
 &  & 10 & 4\\
 &  & $\geq11$ & 0\\
 \hline
    \end{tabular}
    \begin{tabular}[t]{rrrr}
    $n_d$ & $D$ & $N$ & \#terms\\
    \hline\hline
16 & 5 & 2 & 121\\
 &  & 3 & 2745\\
 &  & 4 & 7413\\
 &  & 5 & 7469\\
 &  & 6 & 4103\\
 &  & 7 & 1460\\
 &  & 8 & 390\\
 &  & 9 & 82\\
 &  & 10 & 10\\
 &  & $\geq11$ & 0\\
 & 6 & 2 & 125\\
 &  & 3 & 2786\\
 &  & 4 & 7511\\
 &  & 5 & 7595\\
 &  & 6 & 4196\\
 &  & 7 & 1497\\
 &  & 8 & 401\\
 &  & 9 & 86\\
 &  & 10 & 11\\
 &  & $\geq11$ & 0\\
 \hline
    \end{tabular}
    \begin{tabular}[t]{rrrr}
    $n_d$ & $D$ & $N$ & \#terms\\
    \hline\hline
16 & 7 & 2 & 125\\
 &  & 3 & 2789\\
 &  & 4 & 7516\\
 &  & 5 & 7599\\
 &  & 6 & 4197\\
 &  & 7 & 1497\\
 &  & 8 & 401\\
 &  & 9 & 86\\
 &  & 10 & 11\\
 &  & $\geq11$ & 0\\
 & $\geq8$ & 2 & 126\\
 &  & 3 & 2789\\
 &  & 4 & 7516\\
 &  & 5 & 7599\\
 &  & 6 & 4197\\
 &  & 7 & 1497\\
 &  & 8 & 401\\
 &  & 9 & 86\\
 &  & 10 & 11\\
 &  & $\geq11$ & 0\\
    \hline
    \end{tabular}
    \caption{Results for type 3 terms for $n_d=16$, obtained with the Hilbert series method. For an empty column take the value above it. $n_d$ must be even.}
    \label{tab:resultstype3c}
\end{table}

\begin{table}
    \centering
      \begin{tabular}[t]{rrrr}
    $n_d$ & $D$ & $N$ &\#terms\\
    \hline\hline
13 & 3 & 2 & 5\\
 &  & 3 & 130\\
 &  & 4 & 297\\
 &  & 5 & 204\\
 &  & 6 & 53\\
 &  & 7 & 7\\
 &  & 8 & 2\\
 &  & 9 & 1\\
 &  & $\geq10$ & 0\\
 & 5 & 2 & 1\\
 &  & 3 & 38\\
 &  & 4 & 105\\
 &  & 5 & 154\\
 &  & 6 & 160\\
 &  & 7 & 76\\
 &  & 8 & 10\\
 &  & $\geq9$ & 0\\
 & 7 & $\leq3$ & 0\\
 &  & 4 & 2\\
 &  & 5 & 9\\
 &  & 6 & 8\\
 &  & 7 & 13\\
 &  & 8 & 22\\
 &  & 9 & 10\\
 &  & $\geq10$ & 0\\
 & 9 & $\leq6$ & 0\\
 &  & 7 & 1\\
 &  & 8 & 0\\
 &  & 9 & 0\\
 &  & 10 & 4\\
 &  & 11 & 2\\
 &  & 12 & 0\\
 & 11 & $\leq11$ & 0\\
 &  & 12 & 1\\
     \hline
    \end{tabular}
    \begin{tabular}[t]{rrrr}
    $n_d$ & $D$ & $N$ &\#terms\\
    \hline\hline
14 & 2 & 2 & 3\\
 &  & 3 & 59\\
 &  & 4 & 70\\
 &  & 5 & 23\\
 &  & 6 & 4\\
 &  & 7 & 1\\
 &  & $\geq8$ & 0\\
 & 4 & 2 & 11\\
 &  & 3 & 268\\
 &  & 4 & 779\\
 &  & 5 & 872\\
 &  & 6 & 465\\
 &  & 7 & 116\\
 &  & 8 & 15\\
 &  & 9 & 3\\
 &  & 10 & 1\\
 &  & $\geq11$ & 0\\
 & 6 & 2 & 0\\
 &  & 3 & 14\\
 &  & 4 & 94\\
 &  & 5 & 148\\
 &  & 6 & 173\\
 &  & 7 & 166\\
 &  & 8 & 77\\
 &  & 9 & 10\\
 &  & $\geq10$ & 0\\
 \hline
    \end{tabular}
    \begin{tabular}[t]{rrrr}
    $n_d$ & $D$ & $N$ &\#terms\\
    \hline\hline
14 & 8 & $\leq4$ & 0\\
 &  & 5 & 2\\
 &  & 6 & 9\\
 &  & 7 & 8\\
 &  & 8 & 13\\
 &  & 9 & 22\\
 &  & 10 & 10\\
 &  & $\geq11$ & 0\\
 & 10 & $\leq7$ & 0\\
 &  & 8 & 1\\
 &  & 9 & 0\\
 &  & 10 & 0\\
 &  & 11 & 4\\
 &  & 12 & 2\\
 & $\geq12$ & all & 0\\
 \hline
    \end{tabular}
\caption{Results for the number of terms of type 4 for $n_d=13,14$, obtained with the Hilbert series method. For an empty column take the value above it. $n_d-D$ must be even.}
    \label{tab:resultstype4c}
\end{table}

\begin{table}
    \centering
      \begin{tabular}[t]{rrrr}
    $n_d$ & $D$ & $N$ &\#terms\\
    \hline\hline
15 & 3 & 2 & 21\\
 &  & 3 & 579\\
 &  & 4 & 1396\\
 &  & 5 & 1095\\
 &  & 6 & 389\\
 &  & 7 & 82\\
 &  & 8 & 17\\
 &  & 9 & 2\\
 &  & $\geq10$ & 0\\
 & 5 & 2 & 5\\
 &  & 3 & 310\\
 &  & 4 & 1137\\
 &  & 5 & 1616\\
 &  & 6 & 1351\\
 &  & 7 & 631\\
 &  & 8 & 142\\
 &  & 9 & 16\\
 &  & 10 & 3\\
 &  & 11 & 1\\
 &  & 12 & 0\\
 \hline
    \end{tabular}
    \begin{tabular}[t]{rrrr}
    $n_d$ & $D$ & $N$ &\#terms\\
    \hline\hline
15 & 7 & $\leq3$ & 0\\
 &  & 4 & 42\\
 &  & 5 & 123\\
 &  & 6 & 157\\
 &  & 7 & 175\\
 &  & 8 & 168\\
 &  & 9 & 78\\
 &  & 10 & 10\\
 &  & $\geq11$ & 0\\
 & 9 & $\leq5$ & 0\\
 &  & 6 & 2\\
 &  & 7 & 9\\
 &  & 8 & 8\\
 &  & 9 & 13\\
 &  & 10 & 22\\
 &  & 11 & 10\\
 &  & 12 & 0\\
 & 11 & $\leq8$ & 0\\
 &  & 9 & 1\\
 &  & 10 & 0\\
 &  & 11 & 0\\
 &  & 12 & 4\\
 \hline
    \end{tabular}
\caption{Results for the number of terms of type 4 for $n_d=15$, obtained with the Hilbert series method. For an empty column take the value above it. $n_d-D$ must be even.}
    \label{tab:resultstype4d}
\end{table}

\begin{table}
    \centering
      \begin{tabular}[t]{rrrr}
    $n_d$ & $D$ & $N$ &\#terms\\
    \hline\hline
16 & 2 & 2 & 10\\
 &  & 3 & 142\\
 &  & 4 & 178\\
 &  & 5 & 73\\
 &  & 6 & 19\\
 &  & 7 & 5\\
 &  & $\geq8$ & 0\\
 & 4 & 2 & 44\\
 &  & 3 & 1688\\
 &  & 4 & 5426\\
 &  & 5 & 6227\\
 &  & 6 & 3648\\
 &  & 7 & 1174\\
 &  & 8 & 226\\
 &  & 9 & 36\\
 &  & 10 & 5\\
 &  & $\geq11$ & 0\\
 & 6 & 2 & 0\\
 &  & 3 & 175\\
 &  & 4 & 1126\\
 &  & 5 & 1987\\
 &  & 6 & 2102\\
 &  & 7 & 1537\\
 &  & 8 & 676\\
 &  & 9 & 148\\
 &  & 10 & 16\\
 &  & 11 & 3\\
 &  & 12 & 1\\
 \hline
    \end{tabular}
    \begin{tabular}[t]{rrrr}
    $n_d$ & $D$ & $N$ &\#terms\\
    \hline\hline
16 & 8 & $\leq3$ & 0\\
 &  & 4 & 4\\
 &  & 5 & 52\\
 &  & 6 & 128\\
 &  & 7 & 158\\
 &  & 8 & 175\\
 &  & 9 & 168\\
 &  & 10 & 78\\
 &  & 11 & 10\\
 &  & 12 & 0\\
 & 10 & $\leq6$ & 0\\
 &  & 7 & 2\\
 &  & 8 & 9\\
 &  & 9 & 8\\
 &  & 10 & 13\\
 &  & 11 & 22\\
 &  & 12 & 10\\
 & 12 & $\leq9$ & 0\\
 &  & 10 & 1\\
 &  & $\geq11$ & 0\\
    \hline
    \end{tabular}
\caption{Results for the number of terms of type 4 for $n_d=16$, obtained with the Hilbert series method. For an empty column take the value above it. $n_d-D$ must be even.}
    \label{tab:resultstype4e}
\end{table}

\section{Explicit Hilbert series}\label{app:explicit_Hilbert}

The operators of types 1-4 obtained by the Hilbert series, are given by:
{\footnotesize
\begingroup
\allowdisplaybreaks
%
\endgroup
}

\bibliographystyle{JHEP}
\bibliography{references}

\providecommand{\href}[2]{#2}\begingroup\raggedright\begin{thebibliography}{10}

\bibitem{Weinberg:1968de}
S.~Weinberg, {\it {Nonlinear realizations of chiral symmetry}},  {\em Phys.
  Rev.} {\bf 166} (1968) 1568--1577.

\bibitem{Weinberg:1978kz}
S.~Weinberg, {\it {Phenomenological Lagrangians}},  {\em Physica A} {\bf 96}
  (1979), no.~1-2 327--340.

\bibitem{Gasser:1983yg}
J.~Gasser and H.~Leutwyler, {\it {Chiral Perturbation Theory to One Loop}},
  {\em Annals Phys.} {\bf 158} (1984) 142.

\bibitem{Gasser:1984gg}
J.~Gasser and H.~Leutwyler, {\it {Chiral Perturbation Theory: Expansions in the
  Mass of the Strange Quark}},  {\em Nucl. Phys. B} {\bf 250} (1985) 465--516.

\bibitem{Weinberg:1979sa}
S.~Weinberg, {\it {Baryon and Lepton Nonconserving Processes}},  {\em Phys.
  Rev. Lett.} {\bf 43} (1979) 1566--1570.

\bibitem{Buchmuller:1985jz}
W.~Buchmuller and D.~Wyler, {\it {Effective Lagrangian Analysis of New
  Interactions and Flavor Conservation}},  {\em Nucl. Phys. B} {\bf 268} (1986)
  621--653.

\bibitem{Grzadkowski:2010es}
B.~Grzadkowski, M.~Iskrzynski, M.~Misiak, and J.~Rosiek, {\it {Dimension-Six
  Terms in the Standard Model Lagrangian}},  {\em JHEP} {\bf 10} (2010) 085,
  [\href{http://arxiv.org/abs/1008.4884}{{\tt arXiv:1008.4884}}].

\bibitem{Scherer:1994wi}
S.~Scherer and H.~W. Fearing, {\it {Field transformations and the classical
  equation of motion in chiral perturbation theory}},  {\em Phys. Rev. D} {\bf
  52} (1995) 6445--6450, [\href{http://arxiv.org/abs/hep-ph/9408298}{{\tt
  hep-ph/9408298}}].

\bibitem{Bijnens:1999sh}
J.~Bijnens, G.~Colangelo, and G.~Ecker, {\it {The Mesonic chiral Lagrangian of
  order $p^6$}},  {\em JHEP} {\bf 02} (1999) 020,
  [\href{http://arxiv.org/abs/hep-ph/9902437}{{\tt hep-ph/9902437}}].

\bibitem{Grosse-Knetter:1993tae}
C.~Grosse-Knetter, {\it {Effective Lagrangians with higher derivatives and
  equations of motion}},  {\em Phys. Rev. D} {\bf 49} (1994) 6709--6719,
  [\href{http://arxiv.org/abs/hep-ph/9306321}{{\tt hep-ph/9306321}}].

\bibitem{Fearing:1994ga}
H.~W. Fearing and S.~Scherer, {\it {Extension of the chiral perturbation theory
  meson Lagrangian to order $p^6$}},  {\em Phys. Rev. D} {\bf 53} (1996)
  315--348, [\href{http://arxiv.org/abs/hep-ph/9408346}{{\tt hep-ph/9408346}}].

\bibitem{Bijnens:2018lez}
J.~Bijnens, N.~Hermansson-Truedsson, and S.~Wang, {\it {The order p$^{8}$
  mesonic chiral Lagrangian}},  {\em JHEP} {\bf 01} (2019) 102,
  [\href{http://arxiv.org/abs/1810.06834}{{\tt arXiv:1810.06834}}].

\bibitem{Bijnens:2001bb}
J.~Bijnens, L.~Girlanda, and P.~Talavera, {\it {The Anomalous chiral Lagrangian
  of order $p^6$}},  {\em Eur. Phys. J. C} {\bf 23} (2002) 539--544,
  [\href{http://arxiv.org/abs/hep-ph/0110400}{{\tt hep-ph/0110400}}].

\bibitem{Ebertshauser:2001nj}
T.~Ebertshauser, H.~W. Fearing, and S.~Scherer, {\it {The Anomalous chiral
  perturbation theory meson Lagrangian to order $p^6$ revisited}},  {\em Phys.
  Rev. D} {\bf 65} (2002) 054033,
  [\href{http://arxiv.org/abs/hep-ph/0110261}{{\tt hep-ph/0110261}}].

\bibitem{Benvenuti:2006qr}
S.~Benvenuti, B.~Feng, A.~Hanany, and Y.-H. He, {\it {Counting BPS Operators in
  Gauge Theories: Quivers, Syzygies and Plethystics}},  {\em JHEP} {\bf 11}
  (2007) 050, [\href{http://arxiv.org/abs/hep-th/0608050}{{\tt
  hep-th/0608050}}].

\bibitem{Feng:2007ur}
B.~Feng, A.~Hanany, and Y.-H. He, {\it {Counting gauge invariants: The
  Plethystic program}},  {\em JHEP} {\bf 03} (2007) 090,
  [\href{http://arxiv.org/abs/hep-th/0701063}{{\tt hep-th/0701063}}].

\bibitem{Gray:2008yu}
J.~Gray, A.~Hanany, Y.-H. He, V.~Jejjala, and N.~Mekareeya, {\it {SQCD: A
  Geometric Apercu}},  {\em JHEP} {\bf 05} (2008) 099,
  [\href{http://arxiv.org/abs/0803.4257}{{\tt arXiv:0803.4257}}].

\bibitem{Jenkins:2009dy}
E.~E. Jenkins and A.~V. Manohar, {\it {Algebraic Structure of Lepton and Quark
  Flavor Invariants and CP Violation}},  {\em JHEP} {\bf 10} (2009) 094,
  [\href{http://arxiv.org/abs/0907.4763}{{\tt arXiv:0907.4763}}].

\bibitem{Hanany:2010vu}
A.~Hanany, E.~E. Jenkins, A.~V. Manohar, and G.~Torri, {\it {Hilbert Series for
  Flavor Invariants of the Standard Model}},  {\em JHEP} {\bf 03} (2011) 096,
  [\href{http://arxiv.org/abs/1010.3161}{{\tt arXiv:1010.3161}}].

\bibitem{Lehman:2015via}
L.~Lehman and A.~Martin, {\it {Hilbert Series for Constructing Lagrangians:
  expanding the phenomenologist's toolbox}},  {\em Phys. Rev. D} {\bf 91}
  (2015) 105014, [\href{http://arxiv.org/abs/1503.07537}{{\tt
  arXiv:1503.07537}}].

\bibitem{Henning:2015daa}
B.~Henning, X.~Lu, T.~Melia, and H.~Murayama, {\it {Hilbert series and operator
  bases with derivatives in effective field theories}},  {\em Commun. Math.
  Phys.} {\bf 347} (2016), no.~2 363--388,
  [\href{http://arxiv.org/abs/1507.07240}{{\tt arXiv:1507.07240}}].

\bibitem{Lehman:2015coa}
L.~Lehman and A.~Martin, {\it {Low-derivative operators of the Standard Model
  effective field theory via Hilbert series methods}},  {\em JHEP} {\bf 02}
  (2016) 081, [\href{http://arxiv.org/abs/1510.00372}{{\tt arXiv:1510.00372}}].

\bibitem{Henning:2015alf}
B.~Henning, X.~Lu, T.~Melia, and H.~Murayama, {\it {2, 84, 30, 993, 560, 15456,
  11962, 261485, ...: Higher dimension operators in the SM EFT}},  {\em JHEP}
  {\bf 08} (2017) 016, [\href{http://arxiv.org/abs/1512.03433}{{\tt
  arXiv:1512.03433}}]. [Erratum: JHEP 09, 019 (2019)].

\bibitem{Henning:2017fpj}
B.~Henning, X.~Lu, T.~Melia, and H.~Murayama, {\it {Operator bases,
  $S$-matrices, and their partition functions}},  {\em JHEP} {\bf 10} (2017)
  199, [\href{http://arxiv.org/abs/1706.08520}{{\tt arXiv:1706.08520}}].

\bibitem{Ruhdorfer:2019qmk}
M.~Ruhdorfer, J.~Serra, and A.~Weiler, {\it {Effective Field Theory of Gravity
  to All Orders}},  {\em JHEP} {\bf 05} (2020) 083,
  [\href{http://arxiv.org/abs/1908.08050}{{\tt arXiv:1908.08050}}].

\bibitem{Dujava:2022vqz}
J.~Dujava, {\it {Counting operators in Effective Field Theories}},  bachelor
  thesis, charles university prague, 10, 2022.

\bibitem{Zee:2003mt}
A.~Zee, {\em {Quantum field theory in a nutshell}}.
\newblock Princeton University Press, 2003.

\bibitem{Sakharov:1967dj}
A.~D. Sakharov, {\it {Violation of CP Invariance, C asymmetry, and baryon
  asymmetry of the universe}},  {\em Pisma Zh. Eksp. Teor. Fiz.} {\bf 5} (1967)
  32--35.

\bibitem{Graf:2020yxt}
L.~Graf, B.~Henning, X.~Lu, T.~Melia, and H.~Murayama, {\it {2, 12, 117, 1959,
  45171, 1170086, \textellipsis{}: a Hilbert series for the QCD chiral
  Lagrangian}},  {\em JHEP} {\bf 01} (2021) 142,
  [\href{http://arxiv.org/abs/2009.01239}{{\tt arXiv:2009.01239}}].

\bibitem{Kampf:2013vha}
K.~Kampf, J.~Novotny, and J.~Trnka, {\it {Tree-level Amplitudes in the
  Nonlinear Sigma Model}},  {\em JHEP} {\bf 05} (2013) 032,
  [\href{http://arxiv.org/abs/1304.3048}{{\tt arXiv:1304.3048}}].

\bibitem{Cheung:2014dqa}
C.~Cheung, K.~Kampf, J.~Novotny, and J.~Trnka, {\it {Effective Field Theories
  from Soft Limits of Scattering Amplitudes}},  {\em Phys. Rev. Lett.} {\bf
  114} (2015), no.~22 221602, [\href{http://arxiv.org/abs/1412.4095}{{\tt
  arXiv:1412.4095}}].

\bibitem{Bijnens:2019eze}
J.~Bijnens, K.~Kampf, and M.~Sj\"o, {\it {Higher-order tree-level amplitudes in
  the nonlinear sigma model}},  {\em JHEP} {\bf 11} (2019) 074,
  [\href{http://arxiv.org/abs/1909.13684}{{\tt arXiv:1909.13684}}]. [Erratum:
  JHEP 03, 066 (2021)].

\bibitem{Fonseca:2019yya}
R.~M. Fonseca, {\it {Enumerating the operators of an effective field theory}},
  {\em Phys. Rev. D} {\bf 101} (2020), no.~3 035040,
  [\href{http://arxiv.org/abs/1907.12584}{{\tt arXiv:1907.12584}}].

\bibitem{dai:2020cpk}
L.~Dai, I.~Low, T.~Mehen, and A.~Mohapatra, {\it {Operator Counting and Soft
  Blocks in Chiral Perturbation Theory}},  {\em Phys. Rev. D} {\bf 102} (2020)
  116011, [\href{http://arxiv.org/abs/2009.01819}{{\tt arXiv:2009.01819}}].

\bibitem{Kampf:2021jvf}
K.~Kampf, {\it {The ChPT: top-down and bottom-up}},  {\em JHEP} {\bf 12} (2021)
  140, [\href{http://arxiv.org/abs/2109.11574}{{\tt arXiv:2109.11574}}].

\bibitem{Coleman:1969sm}
S.~R. Coleman, J.~Wess, and B.~Zumino, {\it {Structure of phenomenological
  Lagrangians. 1.}},  {\em Phys. Rev.} {\bf 177} (1969) 2239--2247.

\bibitem{Callan:1969sn}
C.~G. Callan, Jr., S.~R. Coleman, J.~Wess, and B.~Zumino, {\it {Structure of
  phenomenological Lagrangians. 2.}},  {\em Phys. Rev.} {\bf 177} (1969)
  2247--2250.

\bibitem{Gudnason:2021gcp}
S.~B. Gudnason and M.~Nitta, {\it {Reducing the O(3) model as an effective
  field theory}},  {\em JHEP} {\bf 03} (2022) 030,
  [\href{http://arxiv.org/abs/2110.15038}{{\tt arXiv:2110.15038}}].

\bibitem{Bijnens:2009zi}
J.~Bijnens and L.~Carloni, {\it {Leading Logarithms in the Massive O(N)
  Nonlinear Sigma Model}},  {\em Nucl. Phys. B} {\bf 827} (2010) 237--255,
  [\href{http://arxiv.org/abs/0909.5086}{{\tt arXiv:0909.5086}}].

\bibitem{Bijnens:2010xg}
J.~Bijnens and L.~Carloni, {\it {The Massive O(N) Non-linear Sigma Model at
  High Orders}},  {\em Nucl. Phys. B} {\bf 843} (2011) 55--83,
  [\href{http://arxiv.org/abs/1008.3499}{{\tt arXiv:1008.3499}}].

\bibitem{Bijnens:2021hpq}
J.~Bijnens and T.~Husek, {\it {Six-pion amplitude}},  {\em Phys. Rev. D} {\bf
  104} (2021), no.~5 054046, [\href{http://arxiv.org/abs/2107.06291}{{\tt
  arXiv:2107.06291}}].

\bibitem{Manton:2004tk}
N.~S. Manton and P.~Sutcliffe, {\em {Topological solitons}}.
\newblock Cambridge Monographs on Mathematical Physics. Cambridge University
  Press, 2004.

\bibitem{Vermaseren:2000nd}
J.~A.~M. Vermaseren, {\it {New features of FORM}},
  \href{http://arxiv.org/abs/math-ph/0010025}{{\tt math-ph/0010025}}.

\bibitem{Granlund:2016}
T.~Granlund, {\em {{\tt GNU} Multiple Precision Arithmetic Library 6.1.2}},
  2016.

\end{thebibliography}\endgroup
\end{document}